\documentclass[aip,jcp,reprint,longbibliography]{revtex4-2} 
\usepackage{graphicx}
\usepackage{enumitem}
\usepackage{mathtools}
\usepackage{graphicx}   
\usepackage{xcolor}
\usepackage[normalem]{ulem}
\usepackage{caption}
\usepackage{subcaption}
\usepackage{comment}
\usepackage{bbm}
\usepackage{array}
\usepackage{booktabs}
\usepackage{multirow}

\usepackage{amsmath}
\usepackage{amssymb}

\usepackage{booktabs}

\usepackage[breaklinks=true]{hyperref}
\usepackage{breakcites}
\newcommand{\revone}{\textcolor{black}}
\newcommand{\revtwo}{\textcolor{black}}

\begin{document}
\title{An Information Bottleneck Approach for Markov Model Construction}

\author{Dedi Wang}
\thanks{These three authors contributed equally.}
\affiliation{Biophysics Program and Institute for Physical Science and Technology, University of Maryland, College Park, MD 20742, United States}
 
\author{Yunrui Qiu}
\thanks{These three authors contributed equally.}
\affiliation{Department of Chemistry, Theoretical Chemistry Institute, University of Wisconsin-Madison, Madison, WI 53706, United States}
\affiliation{Data Science Institute, University of Wisconsin-Madison, Madison, WI, 53706, United States}

\author{Eric R. Beyerle}
\thanks{These three authors contributed equally.}
\affiliation{Institute for Physical Science and Technology, University of Maryland, College Park, MD 20742, United States}

\author{Xuhui Huang*\footnote{Corresponding author.}}
\email{xhuang@chem.wisc.edu}
\affiliation{Department of Chemistry, Theoretical Chemistry Institute, University of Wisconsin-Madison, Madison, WI 53706, United States}
\affiliation{Data Science Institute, University of Wisconsin-Madison, Madison, WI, 53706, United States}

\author{Pratyush Tiwary*\footnote{Corresponding author.}}
\email{ptiwary@umd.edu}
\affiliation{Department of Chemistry and Biochemistry and Institute for Physical Science and Technology, University of Maryland, College Park, MD 20742, United States}
\affiliation{University of Maryland Institute for Health
Computing, Bethesda, MD 20852, United States}

	\date{\today}
	
\begin{abstract}
\textbf{Abstract\newline }

Markov state models (MSMs) have proven valuable in studying dynamics of protein conformational changes via statistical analysis of molecular dynamics (MD) simulations. In MSMs, the complex configuration space is coarse-grained into conformational states, with dynamics modeled by a series of Markovian transitions among these states at discrete lag times. Constructing the Markovian model at a specific lag time necessitates defining states that circumvent significant internal energy barriers, enabling internal dynamics relaxation within the lag time. This process effectively coarse-grains time and space, integrating out rapid motions within metastable states. Thus, MSMs possess a multi-resolution nature, where the granularity of states can be adjusted according to the time-resolution, offering flexibility in capturing system dynamics. This work introduces a continuous embedding approach for molecular conformations using the state predictive information bottleneck (SPIB), a framework that unifies dimensionality reduction and state space partitioning via a continuous, machine learned basis set. Without explicit optimization of the VAMP-based scores, SPIB demonstrates state-of-the-art performance in identifying slow dynamical processes and constructing predictive multi-resolution Markovian models. Through applications to well-validated mini-proteins, SPIB showcases unique advantages compared to competing methods. It autonomously and self-consistently adjusts the number of metastable states based on specified minimal time resolution, eliminating the need for manual tuning. While maintaining efficacy in dynamical properties, SPIB excels in accurately distinguishing metastable states and capturing numerous well-populated macrostates. This contrasts with existing VAMP-based methods, which often emphasize slow dynamics at the expense of incorporating numerous sparsely populated states. Furthermore, SPIB's ability to learn a low-dimensional continuous embedding of the underlying MSMs enhances the interpretation of dynamic pathways. With these benefits, we propose SPIB as an easy-to-implement methodology for end-to-end MSMs construction.
\end{abstract}

	\maketitle

\section{Introduction}
\label{sec:introduction}

Rapid advances in computational power have made molecular dynamics (MD) a powerful tool for studying molecular systems\cite{Shaw2021}. By implementing physical laws in a simulation, MD enables us to track the time evolution of generic molecular systems in an all-atom, femtosecond resolution\cite{Frenkel2001}. However, all-atom MD simulations commonly face challenges in capturing the long-timescale dynamics of molecular processes. MD trajectories are typically complex, high-dimensional time-sequence data, making comprehension challenging. Consequently, how to accurately model the long-timescale dynamics, and how to make the deluge of data generated from MD simulations understandable to humans are still open questions.\cite{konovalov2021markov, Wang2020} 

A powerful and popular analysis method to study the dynamical behavior of any molecular system displaying a sufficient level of complexity is through Markov state models (MSMs)\cite{Swope2004a,Noe2007,Chodera2007}. MSMs can not only predict long-timescale dynamics using multiple short MD trajectories but also serve as a bridge between high-resolution MD simulations and a more macroscopic description of the dynamics.\cite{konovalov2021markov, prinz2011markov, Wang2018} This makes them a powerful tool for understanding kinetic processes in complex molecular systems.

The conventional pipeline for constructing MSMs involves several intricate steps, such as featurization, dimension reduction, clustering, and kinetic lumping.\cite{konovalov2021markov, Wang2018, Mardt2018, Nagel2023a, yik2023step, Kolloff2024, unarta2021role, prinz2011markov} During featurization, MD conformations are typically aligned or transformed into internal coordinate features to eliminate the translation and rotation of the entire molecules. Subsequently, dimensionality reduction is usually performed based on these features to identify low-dimensional, information-dense collective variables (CVs). The MD conformations are then projected onto the CV space and further grouped into hundreds or thousands of microstates. For better interpretation, microstates can be lumped into a few metastable macrostates based on their kinetic similarities, and the MD trajectories can be coarse-grained into time-series sequences of state indexes. By choosing an appropriate lag time, the transitions between states can be modeled using Markovian jump processes.

With the development of various algorithms advancing the field of MSMs, it is worth noting that the specific construction of MSMs and selection of hyperparameters can substantially impact the quality of the final kinetic model. To quantitatively evaluate the performance of different MSMs and thereby facilitate their construction, a number of metrics have been developed. These metrics encompass various model aspects, including enhancing the metastability of macrostates in the kinetic model\cite{Chodera2007, sheong2015automatic}, optimizing approximations of the principle eigenmodes of dynamics\cite{gu2022rpnet}, and maximizing the model's capability to capture the leading slowest dynamics\cite{Noe2013, Mardt2018}. The third approach, known as the variational approach for Markov processes (VAMP), has now become the most popular. It offers a variational score to measure the difference between the eigenmodes approximated by the model and those of the true dynamical propagator, serving as an objective function to optimize the models.

A number of approaches leveraging artificial neural networks have demonstrated efficacy within the context of variational optimization workflows. Recently, VAMPnets, which employ the VAMP and Koopman theory as their guiding principle, have been developed as a single end-to-end data-driven model to directly map molecular coordinates to coarse-grained macrostates.\cite{Mardt2018} Later, a state-free reversible VAMPnets (SRVs) is specifically introduced to learn nonlinear approximates to the leading slow eigenfunctions\cite{Chen2019} and has also been generalized to biased simulations\cite{bonati2021,shmilovich2023}. Time-lagged autoencoders utilize auto-associative neural networks to reconstruct a time-lagged signal, and embedded variables extracted from the bottleneck layer can serve as the CVs.\cite{Wehmeyer2018} Another approach, variational dynamics encoders (VDEs), combine time-lagged reconstruction loss and autocorrelation maximization within a variational autoencoder to approximate the dynamical propagator.\cite{Hernandez2018}

In this work, we demonstrate a robust protocol for MSMs construction through the use of an information bottleneck approach called State Predictive Information Bottleneck (SPIB).\cite{Wang2021} In previous studies, SPIB learns useful low dimensional CVs for enhanced sampling in molecular simulations to speed up the diverse processes ranging from permeation\cite{Mehdi2022} and dissociation\cite{Beyerle2024,lee2024calculating} of medically relevant ligands, conformational changes in proteins\cite{Vani2023a,Vani2023b} to nucleation of crystal polymorphs\cite{Zou2023,Wang2023}. Here, we focus on demonstrating SPIB as a state-of-the-art approach for automatic construction of multi-resolution MSMs. Different from the existing methods, such as VAMP-based methodologies, which maximizes the Rayleigh coefficient or VAMP score, SPIB integrates the information bottleneck framework with a simple heuristic of the state metastability at a pre-specified lag time to achieve feature extraction and state division in a unified approach. By leveraging this lag time parameter, SPIB facilitates adaptive adjustment of the number of metastable states within the final kinetic model, enabling the automatic generation of MSMs at different resolutions. As a result, SPIB facilitates the automatic clustering and projection of the MD simulation data into a few macrostates and learns directly from MD trajectories a low dimensional latent space where these states are cleanly separated into metastable states, providing a single end-to-end framework that integrates dimension reduction, clustering, and lumping tasks. 

As tests for implementation, we utilize long folding-unfolding MD trajectories of three small proteins (Trp-cage\cite{lindorff2011fast}, villin headpiece [HP35]\cite{piana2012protein}, and WW domain\cite{lindorff2011fast}) from D. E. Shaw
Research (DESRES). Specifically, we compare SPIB to other state-of-the-art methods including VAMPnets\cite{Mardt2018}, traditional MSMs coarse-grained with robust Perron cluster-cluster analysis (PCCA+) algorithm\cite{Deuflhard2005} or the most probable path (MPP) algorithm\cite{Jain2012}  on a set of principal components (PCs), and traditional MSMs coarse-grained with PCCA+ or MPP on a set of time-lagged independent components (tICs).\cite{noe2015kinetic} 
We evaluate the MSMs constructed from different methods using a set of metrics that quantify both the dynamic and static performances of the models. Through this extensive evaluation, we find that SPIB is competitive with or superior to other state-of-the-art methods for creation of MSMs while offering, in addition, a continuous latent space for interpretation of the kinetics and dynamics. Moreover, SPIB exhibits robustness in hyperparameter value selection and demonstrates reduced susceptibility to overfitting, further solidifying its efficacy and reliability in modeling complex systems.

\section{Methods}
\label{sec:methods}
\subsection{Markov State Model (MSM)}
Markov state models (MSMs) are discrete-state and discrete-time kinetic models that enable the approximation of the Markovian dynamical propagator governing the molecular kinetics across the full configuration space. MSMs can be used to coarse grain MD configurations into easily comprehensible states. They can also be used to integrate a large ensemble of short trajectories to predict long-term kinetic and equilibrium thermodynamic properties.\cite{cao2023integrative, qiu2023efficient, liu2022kinetic, yik2023step, liu2023graphvamp, husic2018markov, prinz2011markov, konovalov2021markov, Wang2018, voelz2010molecular,unarta2021role} Popularized for biological systems in the mid-2000s\cite{Swope2004a, Swope2004b,Noe2007, voelz2010molecular}, the use of MSMs has exploded in the last decade due to publication of easy-to-use code\cite{Scherer2015,Harrigan2017} and further developments in the implementation to increase modeling accuracy \cite{Hoffmann2022,cao2023integrative, cao2020advantages, dominic2023memory}. Here we summarize the key ideas underlying MSMs.

Let $\mathbf{X}\subseteq\mathbb{R}^{3N}$ denote a high-dimensional signal characterizing the configuration of a generic molecular system. Then a MSM is constructed by clustering $\mathbf{X}$ into a set of states $\mathbf{y}=\left\{1, 2, 3,\ldots, S\right\}$. For the folding of mini-proteins, $\mathbf{X}$ could be expressed in terms of internal coordinates, such as the minimal residue-residue distances. The states $\mathbf{y}$ could comprise the folded state, unfolded state, misfolded state and some important intermediate states. Let the values measured at time $t$ be denoted by $\mathbf{X}_t$ and $\mathbf{y}_t$ respectively. In this way, the trajectory $\{\mathbf{X}_t\}_{t=0}^{T}$ is coarse-grained to a discrete trajectory $\{\mathbf{y}_{t}\}_{t=0}^{T}$ that indicates which state the system visits at time t. Using statistics from $\mathbf{y}_t$, we can calculate the conditional probabilities of moving between each of the discrete states between a user-defined lag time $\tau$ and store the conditional probabilities in a transition probability matrix (TPM) $\mathbf{T}(\tau)$. If the model is validated as Markovian, the long-timescale dynamics could be modeled through the first-order master equation: $\mathbf{T}(n\tau)=\mathbf{T}^{n}(\tau)$. Both  thermodynamic properties (e.g. stationary state populations) and kinetic (e.g mean first passage time (MFPT), transition pathways and implied timescales (ITS)) can then be calculated from the TPM. Specifically, if the TPM is diagonalized with the left and right eigenvectors and eigenvalues: $T_{ij}(\tau)=\phi^T_i(\tau)\lambda_i(\tau)\psi_j(\tau)$, the stationary populations can be directly obtained from the left eigenvector $\phi_{1}(\tau)$ (if the TPM is row-normalized) with eigenvalues equal to 1. The remaining eigenvectors can be utilized to infer the transition mechanisms between the states. And the ITS for different eigenmodes can be defined as $t_i(\tau)=\frac{-\tau}{\ln(\lambda_i(\tau))}$. We refer to Refs. \onlinecite{konovalov2021markov}, \onlinecite{prinz2011markov}, \onlinecite{Kolloff2024} for an excellent recent technical review of MSMs.

\subsection{Variational approach for Markov processes (VAMP)}

The variational approach for Markov processes (VAMP), developed by Wu and Noé \cite{ wu2017variational, wu2020variational}, provides a variational framework to approximate the dynamical propagator of general Markovian dynamical processes, whether reversible or irreversible, stationary or non-stationary. The VAMP theory introduces the VAMP-r score, defined as the sum of the singular values raised to the power $r$ of the approximated propagator, to quantify the difference between the approximated leading singular functions of the dynamical propagator and the ground truth. A higher score indicates a superior approximation of the singular functions which represent the leading slowest dynamical modes of the system. Based on this, VAMPnets have been developed\cite{Mardt2018}, emerging as an end-to-end, neural-network-based, unsupervised method for constructing MSMs. VAMPnets can directly map internal coordinate features to macrostate assignments probabilities, thereby replacing the intricate intermediate pipeline. Additionally, if the studied dynamics are time-reversible and detailed balanced, a specialized variational approach for conformational dynamics (VAC) can be derived.\cite{nuske2014variational, Noe2013}

\subsection{State Predictive Information Bottleneck (SPIB)}

Existing VAMP-based methods typically specialize in either direct state space partitioning, as in VAMPnets\cite{Mardt2018}, or in identifying low-dimensional continuous CVs of the input data, as in the case of SRVs\cite{Chen2019}. In contrast, we utilize an information bottleneck approach for the MSM construction, called State Predictive Information Bottleneck (SPIB).\cite{Wang2021} This method offers a unified framework that seamlessly integrates both state partitioning and dimension reduction. 

To learn both the number and location of the potential metastable states in the system for constructing MSMs, SPIB is based on a simple heuristic that quantifies metastability. Namely, the central idea is that if one configuration was located at state $i$ at a certain time, then after a short lag time $\Delta t$, it should still have the largest probability to be found at state $i$. This is because if state $i$ is metastable, then its escape time should be much larger than $\Delta t$. Based on this heuristic, in Ref. \onlinecite{Wang2021} we introduced an iterative scheme to learn the number and location of states on-the-fly.

We start with an arbitrary set of state labels for the system $\{\mathbf{y}_{t}\}_{t=0}^{T}$, where both the number and location of labels are some initial guess. The probability that the system starting from $\mathbf{X}$ will be found in state $\mathbf{y}_{\Delta t}=i$ after a lag time $\Delta t$ can be estimated by the following, assuming a stationary distribution:
\begin{equation} 
\begin{aligned} 
\label{eq:state_transition_density}
p(\mathbf{y}_{\Delta t}=i|\mathbf{X})&=\frac{1}{\rho(\mathbf{X})}\lim_{T \to +\infty}\frac{1}{T}\int_{0}^{T} \mathbbm{1}_{\mathbf{y}_{t+\Delta t}=i}\delta(\mathbf{X}-\mathbf{X}_t)dt\\
\text{where}\ &\rho(\mathbf{X})=\lim_{T \to +\infty}\frac{1}{T}\int_{0}^{T}\delta(\mathbf{X}-\mathbf{X}_t)dt.
\end{aligned} 
\end{equation}
Here $\rho(\mathbf{X})$ represents the equilibrium density of $\mathbf{X}$ and $\mathbbm{1}_{\mathbf{y}_{t+\Delta t}=i}$ is the indicator function for state $i$, which is equal to 1 if the trajectory is within state $i$ at time $t+\Delta t$ and equal to 0 otherwise. As it is a function of the input configuration $\mathbf{X}$ and represents a state-transition probability, we call the function $p(\mathbf{y}_{\Delta t}|\mathbf{X})$ as the state-transition density. If the system initiated from a certain high-dimensional configuration $\mathbf{X}$ has the largest probability to be found after lag time $\Delta t$ in some state $i$ from these initial labels, then the label of the configuration $\mathbf{X}$ will be refined and updated to state $i$. Thus, a set of new state labels can be generated by:
\begin{equation} 
\label{eq:label_update}
\hat{\mathbf{y}}_t=\underset{i}{\text{argmax}}\ p(\mathbf{y}_{t+\Delta t}=i|\mathbf{X}_t).
\end{equation}
Based on the new refined state labels, the state-transition density $p(\mathbf{y}_{\Delta t}|\mathbf{X})$ can be re-estimated and the process can be repeated until state labels converge. This refinement in labeling may result in null assignments for certain initial labels, consequently reducing the overall number of states. In this way, starting from arbitrary states, SPIB can learn both the number and location of the potential metastable states in the system dynamically.

Direct estimation of the state-transition density using Eq. \ref{eq:state_transition_density} suffers from the curse of dimensionality as the input feature $\mathbf{X}$ is typically high dimensional for complex systems. To alleviate this problem, SPIB aims to uncover a low dimensional manifold on which the dynamics of the system can be projected for more robust estimation of the state-transition density $p(\mathbf{y}_{\Delta t}|\mathbf{X})$. This provides a unified pipeline for both dimension reduction and state decomposition. Such a low dimensional representation $\mathbf{z}_t$ is assumed to use minimal information from the past signal $\mathbf{X}_t$ to predict its future state label $\mathbf{y}_{t+\Delta t}$ accurately. Such a learning process can be facilitated through the deep variational information bottleneck framework.\cite{Alemi2017,Wang2021} Several modifications have been implemented since the initial publications\cite{Wang2019Past,Wang2021} to enhance the algorithm's robustness, as summarized below.

\begin{figure}[t!]
    \centering
    \includegraphics[width=0.48\textwidth]{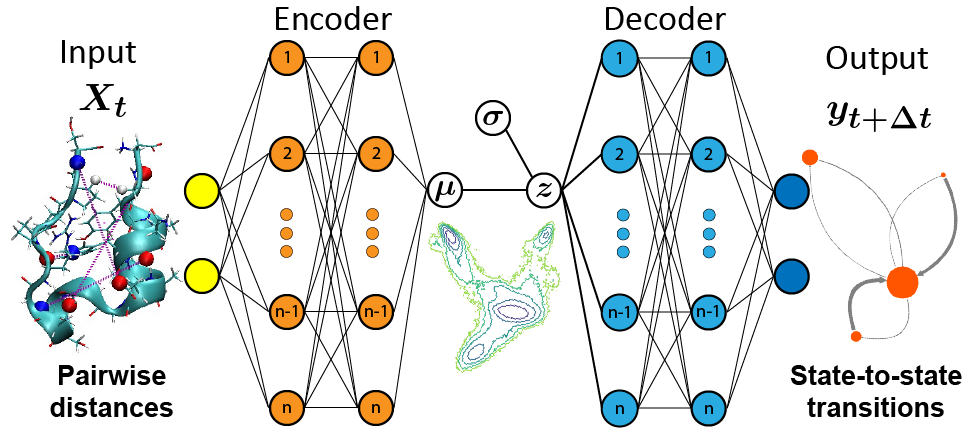}
    \caption{Network architecture employed for SPIB consists of both the encoder and decoder as nonlinear neural networks with two hidden layers. SPIB is designed to take features such as pairwise distances, denoted as input $\mathbf{X}_{t}$, enabling the learning of a low-dimensional latent representation $\mathbf{z}$ for predicting its future state $\mathbf{y}_{t+\Delta t}$ after a lag time $\Delta t$. In this modified architecture, the encoder only outputs the mean $\mathbf{\mu}$, from which the latent representation $\mathbf{z}$ is then sampled utilizing a position-independent trainable standard deviation $\mathbf{\sigma}$. For visualization, the left panel illustrates some minimal residue-residue distances of the Trp-cage system. In the middle, an example of the free energy surface of the learned latent space is displayed. The right panel presents a network plot of the output Markov state model.}
    \label{fig:network}
\end{figure}

The network architecture used for SPIB is shown in Fig. \ref{fig:network}. For a given trajectory $\{\mathbf{X}^1,\cdots,\mathbf{X}^{M}\}$ and its corresponding state labels $\{\mathbf{y}^1,\cdots,\mathbf{y}^{M}\}$, where the length $M$ is sufficiently large, the objective function of SPIB can be formulated as:

\begin{equation} 
\begin{aligned} 
\label{eq:SPIB_obj}
&\underset{\theta}{\text{argmax}}~\mathcal{L}(\theta)=\\
&\frac{1}{M-s}\sum_{n=1}^{M-s} \Bigl[\log q_{\theta}(\mathbf{y}^{n+s}|\mathbf{z}^{n})-\beta \log \frac{p_{\theta}(\mathbf{z}^{n}|\mathbf{X}^n)}{r_{\theta}(\mathbf{z}^{n})} \Bigr]
\end{aligned} 
\end{equation}
where the encoder $p_\theta(\mathbf{z}|\mathbf{X})$, the decoder $q_{\theta}(\mathbf{y}|\mathbf{z})$, and the prior $r_{\theta}(\mathbf{z})$ are probability distributions parameterized by deep neural networks $\theta$. $\mathbf{z}^{n}$ is sampled from $p_{\theta}(\mathbf{z}|\mathbf{X}^n)$ and the time interval between $\mathbf{X}^n$ and $\mathbf{X}^{n+s}$ is the lag time $\Delta t$, or how far into the future SPIB should predict. The first term $\log q_{\theta}(\mathbf{y}^{n+s}|\mathbf{z}^{n})$ measures the ability of our representation to predict the desired target, while the second term $\log \frac{p_{\theta}(\mathbf{z}^{n}|\mathbf{X}^n)}{r_{\theta}(\mathbf{z}^{n})}$ can be interpreted as the complexity penalty that acts as a regulariser. This regularization term encourages the latent space $\mathbf{z}$ to retain less information from the input $\mathbf{X}$, thereby promoting a more compact representation. Such a trade-off between the prediction capacity and model complexity is then controlled by a hyperparameter $\beta\in[0,\infty)$. 

For simplicity, we opt to employ a Gaussian encoder with a constant variance:
\begin{equation} 
\label{eq:gaussian_encoder}
\log p_{\theta}(\mathbf{z}^n|\mathbf{X}^n) = \log \mathcal{N}(\mathbf{z}^n;\mathbf{\mu},\mathbf{\sigma} I)
\end{equation}
where only the mean $\mathbf{\mu}=\mathbf{\mu}_{\theta}(\mathbf{X}^n)$ is the output of a neural network whose input is $\mathbf{X}^n$, while the variance $\mathbf{\sigma}^2$ is a trainable parameter independent of the input $\mathbf{X}$. $I$ is the identity matrix. We fix the variance of $p_{\theta}(\mathbf{z}^n|\mathbf{X}^n)$ to be constant for all $\mathbf{X}$. This enables the learning of a latent space with enhanced homogeneity.

A deep feed forward neural network with softmax outputs is used in the decoder $q_{\theta}(\mathbf{y}|\mathbf{z})$: 
\begin{equation} 
\label{eq:softmax_decoder}
\log q_{\theta}(\mathbf{y}^{n+s}|\mathbf{z}^n) = \sum_{i=1}^{S} y_i^{n+s}\log \mathcal{D}_i(\mathbf{z}^n;\theta)
\end{equation} 
where the state label $\mathbf{y}$ is a one-hot vector of $S$ dimensions and the decoder function $\mathbf{\mathcal{D}}$ is the $S$-dimensional softmax output of a neural network. This use of the softmax output in the decoder allows for fuzzy assignments to the SPIB-predicted states $\mathbf{y}$.

Given that we expect the latent representation $\mathbf{z}$ should demarcate between different metastable states, it is natural to assume a multi-modal distribution for the prior $r_{\theta}(\mathbf{z})$.
Thus, we modify the variational mixture of posteriors prior algorithm\cite{Tomczak2018} to obtain such a multi-modal prior distribution. 
Here, the approximate prior $r_{\theta}(\mathbf{z})$ is a weighted mixture of different posteriors $p_{\theta}(\mathbf{z}|\mathbf{X})$ with some representative-inputs $\{\mathbf{X}_\text{rep}^k\}_{k=1}^K$ in lieu of $\mathbf{X}$:
\begin{equation} 
\label{eq:new_vamp_prior}
r_{\theta}(\mathbf{z}) = \sum_{k=1}^{K}\omega_k\ p_{\theta}(\mathbf{z}|\mathbf{X}_\text{rep}^k),
\end{equation} 
where $K$ is the number of representative-inputs, and $\omega_k$ represents the weight of $p_{\theta}(\mathbf{z}|\mathbf{X}_\text{rep}^k)$ under the constraint $\sum_{k}\omega_k=1$. The algorithm to determine the representative inputs $\{\mathbf{X}_\text{rep}^k\}_{k=1}^K$ operates as follows: Initially, one sample is randomly selected for each initial state to form our initial set of representative inputs $\{\mathbf{X}_\text{rep}^k\}_{k=1}^K$, where $K$ corresponds to the number of initial states. After each iteration of model training and state label refinement, all input samples are mapped to the learned latent space. We compute the center of each newly refined, non-empty metastable state in the learned latent space and identify the nearest sample for each center using Euclidean distance in the latent space. These selected samples then serve as the new representative inputs $\{\mathbf{X}_\text{rep}^k\}_{k=1}^K$ in the subsequent iteration. Consequently, our algorithm automatically adjusts the representative inputs $\{\mathbf{X}_\text{rep}^k\}_{k=1}^K$, and the number of representative inputs $K$ always matches the number of states in $\mathbf{y}$. Moreover, by incorporating the mixture of Gaussians prior to Eq. \ref{eq:SPIB_obj}, the regularization term is able to encourage spatial separation among the state centers within the latent space.

After obtaining the optimal $\theta^*$ by maximizing the SPIB objective function (Eq. \ref{eq:SPIB_obj}), we can revisit Eq. \ref{eq:label_update}. Utilizing the deterministic output of SPIB denoted as $\hat{p}(\mathbf{y}_{t+\Delta t}|\mathbf{X}_t)\equiv \mathbf{\mathcal{D}}(\mathbf{\mu}(\mathbf{X});\Delta t, \theta^*)$, we can effectively approximate the state-transition density $p(\mathbf{y}_{t+\Delta t}|\mathbf{X}_t)$. The rule of state label update can be formulated as:
\begin{equation} 
\label{eq:spib_label_update}
\hat{\mathbf{y}}_t=\underset{i}{\text{argmax}}\ \mathcal{D}_i(\mathbf{\mu}(\mathbf{X}_t);\Delta t, \theta^*).
\end{equation}

The workflow of SPIB is summarized as follows: To initiate the training process, in addition to trajectory data expressed in terms of input features, SPIB requires an initial set of state labels as input. Thus, the initial state labels are generated as the first step. The choice of initial state labels acts as a form of prior information, effectively guiding and shaping the learning process within SPIB. One straightforward approach to generate these initial state labels is to discretize some input order parameters (OPs) based on expert intuition, a method commonly employed in many previous works.\cite{Wang2021,Wang2022,Mehdi2022,Beyerle2022,Zou2023,Wang2023,Vani2023a,Vani2023b,Beyerle2024} However, for more intricate systems, especially those lacking intuitive guidance, such as the protein folding systems explored in this study, the initial assignment of states can be carried out in similar manner as traditional MSM construction. This involves the use of dimension reduction techniques such as PCA or tICA to identify the subset of linearly optimal combination of a large set of input features and commonly used clustering algorithms such as K-means or K-centers to produce the discrete states. Following the generation of initial state labels, the trajectory data $\mathbf{X}$ and state labels $\mathbf{y}$ are input into SPIB. The objective is to find the optimum latent representation which captures the most important features of the past configuration $\mathbf{X}_t$ to predict the future state $\mathbf{y}_{t+\Delta t}$. After this learning process, we can refine the state labels based on Eq. \ref{eq:spib_label_update}, and the new refined state labels are then fed back into SPIB. The processes will be repeated until the converged latent representation and state labels are generated for further analyses. For a clearer understanding of the training process, a pseudo-code outlining the SPIB training procedure is presented in the Supplementary Information (SI).

\subsection{Baseline models and Quantitative Metrics}
SPIB's capability to autonomously discern metastable states within systems suggests it could be a promising tool for MSM construction. To evaluate the quality of the resultant MSM from SPIB, we applied various quantitative metrics and conducted a systematic comparison with MSMs built through different pipelines. Generally, we followed the traditional order of dimensionality reduction, clustering, and lumping, but applied different algorithms for each step. For dimensionality reduction, we used tICA and PCA algorithms; for the clustering step, we keep employing the k-means algorithm; and for the lumping step, we utilized PCCA+\cite{Deuflhard2005} and MPP\cite{Jain2012,Nagel2023a} algorithms. This resulted in four different combinations of pipelines for constructing MSMs. Additionally, to compare with other deep neural network-based methods, we also implemented the commonly used VAMPnets\cite{Mardt2018} as another reference. We note that for both SPIB and VAMPnet, we crisply assign state labels to MD conformations based on the highest output probability of the neural networks. Detailed setups of different methods are outlined in SI.

For quantitative metrics, in line with the benchmark work conducted on the same HP35 trajectory in Refs. \onlinecite{Nagel2023a,Nagel2023b}, and considering traditional score functions, we choose to utilize the generalized matrix Rayleigh quotient (GMRQ)\cite{Husic2016}, metastability score\cite{Chodera2007}, the Shannon entropy\cite{Nagel2023b}, the Davies-Boldin index (DBI)\cite{Nagel2023b} and the implied timescales (ITS). We define these as follows:

\begin{enumerate}
\item GMRQ score\cite{mcgibbon2015variational}: defined as the sum of the top $n$ eigenvalues, $\lambda_i$, of the transition probability matrix (TPM):
\begin{equation}
\text{GMRQ}=\sum_{i=1}^{n}\lambda_i,
\end{equation}
where $n$ is the total number of eigenvalues scored. Based on VAC theory\cite{nuske2014variational}, in cases where the studied dynamics are reversible and detailed-balanced (e.g. protein conformational changes), the sum of the eigenvalues of the approximated propagator (i.e., GMRQ score) can function as a variational score, serving as a lower bound to the ground truth.\cite{Husic2016, mcgibbon2015variational} In these cases, maximizing the VAMP based scores lead to larger eigenvalues for reversible propagators and, consequently, higher GMRQ scores.
Notably, the GMRQ score also functions as a commonly used criterion in cross-validation for determining the optimal hyperparameters of the models while avoiding overfitting. \cite{mcgibbon2015variational}

\item Metastability \revone{$Q$}: defined as the mean of the trace of the TPM, which measures how probable it is for the system to remain in the same state after a lag time $\tau$ parametrizing the TPM. Metastability is a useful metric for evaluating the model, and high metastability typically indicates effective separation of slow inter-state dynamics from fast intra-state dynamics.\cite{Chodera2007} \revone{Explicitly, we define the metastability $Q$ as}
\begin{equation}
Q=\frac{1}{S}\text{tr}\left(\mathbf{T}(\tau)\right),
\end{equation}
\revone{where $S$ is, as previously defined, the number of metastable states spanned by $\mathbf{T}(\tau)$.}

\item Shannon entropy $\text{H}$ of the learned metastable states: defined in the usual information theoretic manner as
\begin{equation}
\text{H}=-\sum_i \pi_i\log(\pi_i),
\end{equation}
with $\pi_i$ the stationary or marginal probability of occupying metastable state $i$. Precisely, $\left(\mathbf{\pi}^T\mathbf{T}(\tau)\right)_{i}=\pi_i$ (the TPM is row-normalized). Hence, a higher value of $\text{H}$ indicates that a significant portion of states is adequately populated, a preference we prioritize over partitionings characterized by a few highly populated states and numerous sparsely populated ones.

\item Additionally, the DBI is the ratio of the average intra-state distance between data points clustered to a single state and the interstate distance between the centers-of-mass of these states and describes how well separated the metastable states are:
\begin{equation}
\text{DBI}=\frac{1}{N}\sum_i\max_{j}\frac{s_i + s_j}{r_{ij}},
\end{equation}
with $s_i$ average distance of a point in state $i$ from the centroid of metastable state $i$ and $r_{ij}$ is the distance between centroids of states $i$ and $j$. A small DBI value indicates well-separated, structurally distinct states. Practically, the DBI is calculated using the implementation in the Scipy python library.\cite{scikit-learn}

\item Implied timescales (ITS): which monitors the timescales, $t_i$, of eigenmode $i$ across different lag time, were calculated for TPMs with different lag time\cite{Swope2004a,Swope2004b}:
\begin{equation}
t_i(\tau)=-\frac{\tau}{\ln|\lambda_i|}.
\end{equation}
where $\tau$ is the lag time used to estimate the TPM, and $\lambda_{i}$ is the $i$-th eigenvalue for the TPM. Typically, if the ITS converge and are independent of the lag time $\tau$, it implies that the dynamics of the model satisfy the first-order master equation: $\lambda(n\tau)= \lambda_i(\tau)^n$. This property could be used to determine the shortest Markovian lag time. The Chapman-Kolmogorov (CK) test can serve as an additional validation tool to examine the Markovian properties of the model.\cite{Bowman2013}

Based on the ITS analysis, two crucial factors can be used to assess the qualities of the MSMs: Markovian lag time and values of converged timescales. The Markovian lag time is the shortest lag time where all the ITS converge and represents the time-resolution for the MSMs. A shorter Markovian lag time indicates better separation of slow inter-state dynamics from fast intra-state dynamics. Additionally, as the lag time is constrained by the trajectory length, constructing an MSM with a shorter Markovian lag time can reduce the demand for simulation length. Furthermore, according to the VAC theory, a model with larger converged timescales demonstrates greater capability to capture the leading slowest dynamics.

\end{enumerate}

\begin{figure}[t!]
    \centering
    \includegraphics[width=0.48\textwidth]{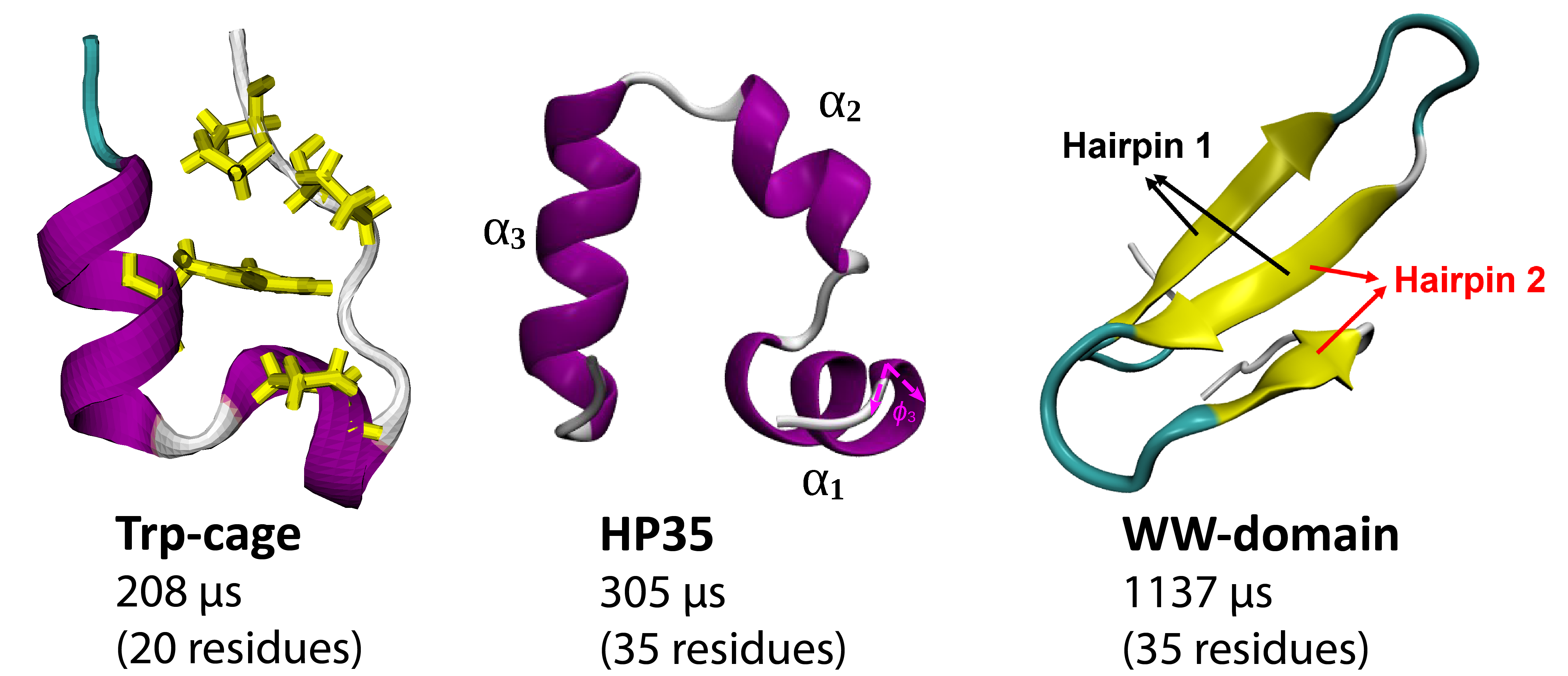}
    \caption{Protein systems investigated in this study. Data for all simulations is obtained from the DESRES protein folding trajectories. The duration of the MD simulation and the number of residues are specified for each case.}
    \label{fig:systems}
\end{figure}

\subsection{Systems, MSM Construction and Validation}
All analyses in this study are performed on the long equilibrium molecular
dynamics trajectories of three mini-proteins from the DESRES group, namely Trp-cage (PDB:2JOF)\cite{lindorff2011fast}, HP35 (PDB:2F4K)\cite{piana2012protein}, and WW domain (PDB:2F21)\cite{lindorff2011fast} (Fig. \ref{fig:systems}). 

All three datasets are featurized with all minimal residue-residue distances (calculated as the closest distance between the heavy atoms of two residues separated in sequence by at least two neighboring residues), resulting in 153 features for Trp-cage, 528 features for both HP35 and WW-domain. \revtwo{The features constructed from distances between the closest heavy atoms offer the advantage of accurately depicting the formation and breaking of local residue contacts, thereby facilitating the identification of structurally well-defined conformational states.} \revtwo{But we note that these embedded features may result in non-differentiable CVs due to the switching of closest atom pairs. Therefore, utilizing distances between $C_\alpha$ atoms or mass centers may yield CVs more appropriate for enhanced sampling.} For computational feasibility, we analyze the HP35 and WW-domain at a resolution 1 ns per frame, but for the smaller Trp-cage system we retain the 0.2 ns per frame resolution of the original trajectory.

For all the algorithms employed in this study, we identified an optimal set of hyperparameters through ten-fold cross-validation using the GMRQ score as the scoring metric (see SI). A more complex model typically possesses greater capacity to capture significant slow dynamics (indicated by a larger GMRQ score). However, due to limited data, a trade-off is necessary between model capacities and generalization abilities. The VAMP-2 score serves as an alternative criterion, similar to the GMRQ score when the dynamics are reversible. 

To prepare the data, we first divide the long equilibrium DESRES trajectory into 100 equal-length short segments. These segments are treated as independent trajectories, mimicking the practical scenario of MSM construction where multiple short trajectories are collected. For \revone{10-fold} cross-validation, the \revone{100} segments are then shuffled and \revone{randomly} subsampled as part of the train-test split procedure for each fold. \revone{This straightforward data processing method has been employed in the previous study\cite{Sidky2019}. Given the sufficiently long protein folding trajectories provided by DESRES, we anticipate that this simple train-test split procedure will adequately ensure that each split represents the dynamics of interest.} All parameters are then scored by considering only the top 3 eigenvalues at a 100 ns MSM lag time for all three systems. The 100ns lag time is validated as Markovian lag time for all three systems based on the ITS analysis and CK test (see SI). All GRMQ scores and metastability values reported in this paper are calculated using MSMs with 100 ns lag time. 

\section{Results}
\label{sec:results}

\subsection{Effect of lag time $\Delta t$ on SPIB}
\begin{figure}[htbp]
     \centering
     \begin{subfigure}[b]{0.45\textwidth}
         \caption{Trp-cage}
         \includegraphics[width=\textwidth]{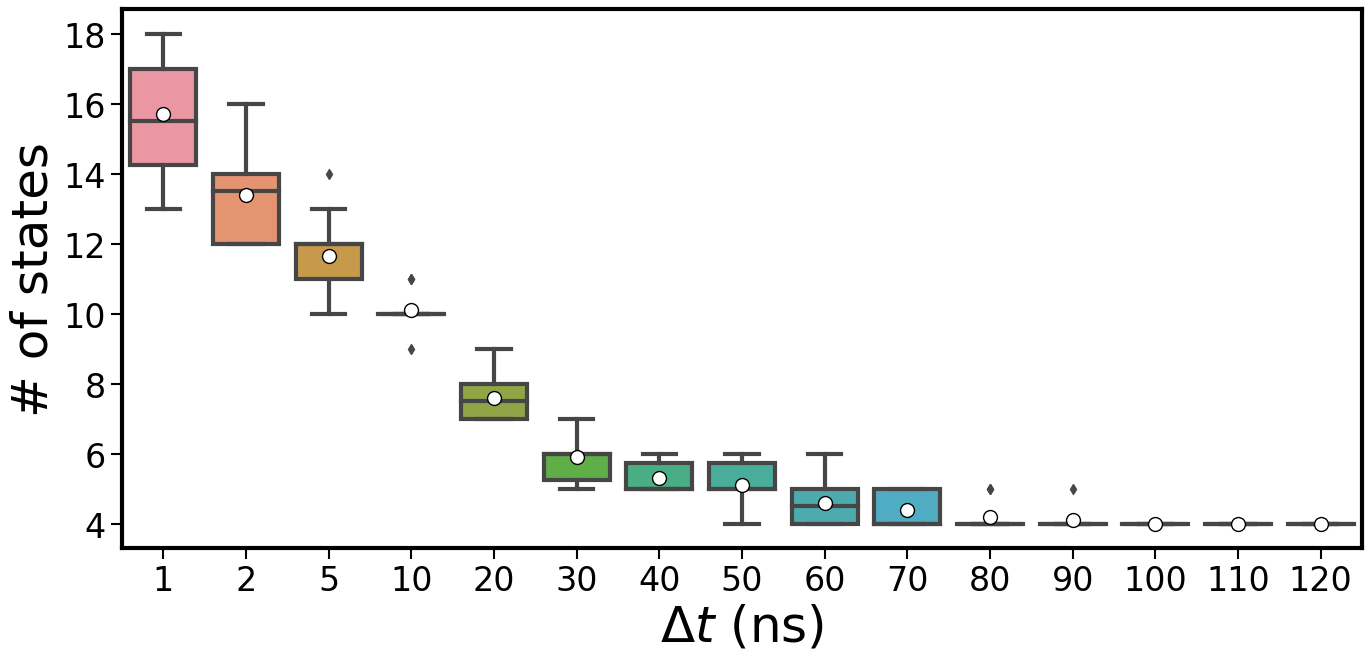}
     \end{subfigure}
     \hfill
     \begin{subfigure}[b]{0.45\textwidth}
         \caption{HP35}
         \includegraphics[width=\textwidth]{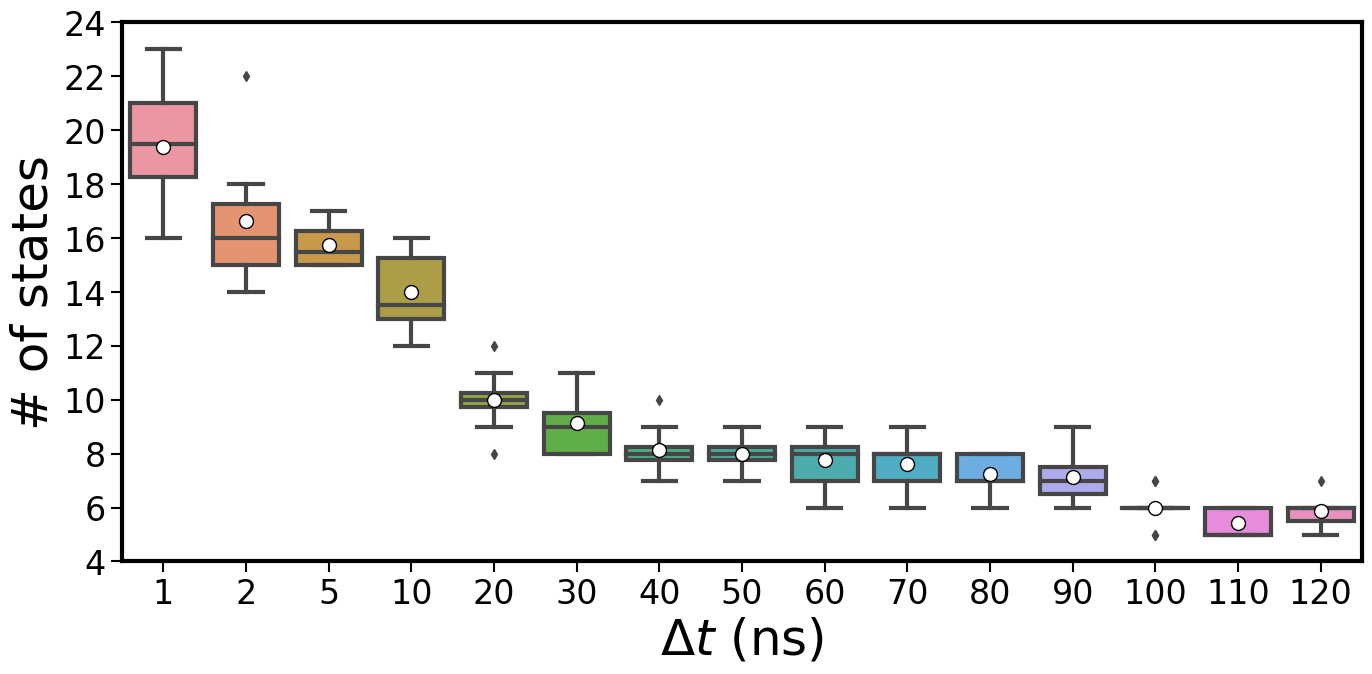}
     \end{subfigure}
     \hfill
     \begin{subfigure}[b]{0.45\textwidth}
         \caption{WW-domain}
         \includegraphics[width=\textwidth]{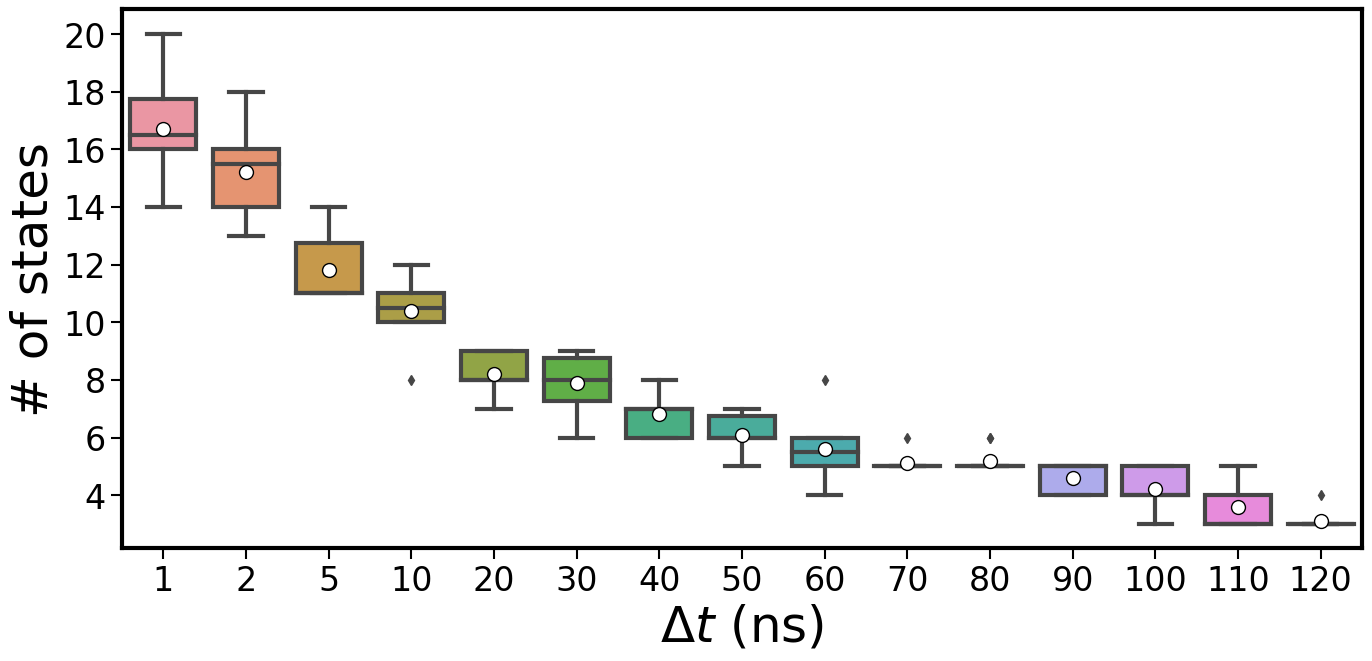}
     \end{subfigure}
     \hfill
        \caption{Impact of different lag time $\Delta t$ choices on the number of converged SPIB states in 10-fold cross-validation for all three systems.}
        \label{fig:spib_dt_n_cv}
\end{figure}

SPIB can be conceptualized as a ``fast mode filter", where the hyperparameter $\Delta t$, representing how far into the future the model should predict, acts as a tool to filter out states with short lifetimes and control the level of dynamic coarse-graining. The choice of the lag time $\Delta t$ plays a crucial role in shaping the simplification of the learning process. When $\Delta t = 0$, SPIB disregards dynamics entirely and focuses solely on clustering the input configuration into distinct states. Conversely, for $\Delta t > 0$, SPIB functions as an effective filter, identifying and excluding dynamics occurring on a timescale faster than $\Delta t$ from consideration. This capability enables the neglect of unnecessary details in the dynamical processes, providing a dynamics-based, coarse-grained understanding. Furthermore, as illustrated in Fig. \ref{fig:spib_dt_n_cv}, the increase in lag time $\Delta t$ leads directly to a decrease in the number of metastable states one expects to find after a delay of $\Delta t$. This highlights an advantage of SPIB, as it automatically adjusts the number of metastable states in the system based on the specified $\Delta t$. As $\Delta t \to \infty $, the number of states will gradually decrease to 1 corresponding to the most stable state of the system.

As Fig. \ref{fig:spib_dt_n_cv} shows, varying $\Delta t$ for SPIB yields different converged numbers of metastable states. This absence of a distinct plateau in the selection of the $\Delta t$ at shorter timescales provides evidence for a rugged free-energy landscape of protein folding. Nevertheless, each choice of different $\Delta t$ and the resulting states are meaningful, adeptly capturing the system's relevant dynamics at the selected temporal resolution and important conformations, as detailed in subsequent subsections, and demonstrating the SPIB method's ability to give detailed insights regarding the hierarchical energy landscapes of simple proteins.

To comprehensively assess the performance of SPIB in both qualitative and quantitative terms, we strategically select two values of $\Delta t$, denoted as ``large $\Delta t$" and ``moderate $\Delta t$". This selection yields two distinct sets of MSMs comprising 4 or 5 states for large $\Delta t$ and approximately 10 states for moderate $\Delta t$. To obtain fewer states for better interpretability, we avoid using $\Delta t$ that is any smaller. Furthermore, as depicted in Fig. \ref{fig:spib_dt_n_cv}, it is evident that a smaller value of $\Delta t$ generally results in a higher variance in the number of states. This variability can be attributed to the presence of states with relatively small populations at short lag times, which may occasionally be overlooked by SPIB. Additionally, the limited sampling and cross-validation procedures further exacerbate the variability observed in the results.

\subsection{Quantitative assessment}

To achieve a more coarse-grained representation of the SPIB states, we tune $\Delta t$ value to yield, on average, 4 or 5 metastable states. Specifically, for Trp-cage, we select $\Delta t = 100$ ns which yielded 4 metastable states using SPIB. For HP35, we use $\Delta t = 110$ ns, which generates models with an average of 5 metastable states. Finally, for the WW-domain, we use $\Delta t = 70$ ns which led to 5 metastable states with SPIB.  We note that, given the results presented in Figure \ref{fig:spib_dt_n_cv}, different lag times within a small range of the selected values for each protein will yield qualitatively similar SPIB models. In this section, we show how these coarse SPIB models can be used as kinetic models competitive or superior to other state-of-the-art methods for MSM construction. These high-performance kinetic models constructed using SPIB are generated without explicitly optimizing along the measured performance metrics and are robust across a wide range of hyperparameter space, as shown in gruesome detail in the SI.

For a fair and meaningful comparison, we ensure that all baseline methods yield the same number of states as obtained by SPIB. This is achieved by adjusting the relevant hyperparameters: the metastability criterion $Q_\text{min}$ for MPP, the number of clusters $m$ for PCCA+, and the number of output states for VAMPnets. Other hyperparameters are fine-tuned through cross-validation using the GMRQ score to ensure that the comparison is conducted among equivalently optimized MSMs constructed by different methods (see details in the SI).  The comprehensive results are summarized in Table \ref{tab:large_dt_quantitative_comparison}, where we see how the performance of SPIB compares to the state-of-the-art methods of PCA or tICA plus PCCA+ or MPP and VAMPnet. 

\revtwo{For the SPIB models examined, the dimensionality of the SPIB latent space used to construct MSMs is always two-dimensional. For the PCA and tICA approaches, the GMRQ is maximized when using a higher dimensionality. For PCA, we utilized the 10, 10, and 20 PCs with the highest variances for constructing MSMs for Trp-cage, HP35, and WW-domain, respectively. For Trp-cage, HP35, and WW-domain, we use 3, 4, and 6 tICA components for MSM construction, respectively. The GMRQ scores for MSMs constructed on PCA, tICA, and SPIB spaces of varying dimensionality are given in the SI. }

Upon analyzing the results, it becomes evident that VAMPnet consistently produces the highest GMRQ scores and metastability in the training data across all three systems since maximizing the VAMP-2 score for reversible dynamics will, in turn, maximize the eigenvalues of the TPM, resulting in larger GMRQ scores and generally higher metastabilities. However, without employing the VAMP-based score as the objective function, which excels in maximizing the GMRQ score, SPIB demonstrates comparable performance to tICA-PCCA+ and VAMPnet regarding the GMRQ score and metastability in the validation data. Beyond exhibiting similar proficiency to these state-of-the-art methods in capturing slow dynamics, SPIB also consistently attains well-populated and structurally distinct states, leading to comparable entropy and DBI across all three systems when compared to tICA-PCCA+ and VAMPnet. 

It's worth noting that other methods occasionally achieve even higher entropy, as observed with PCA-PCCA+ in Trp-cage and PCA-MPP in WW-domain, or lower DBI, such as PCA-PCCA+ in WW-domain. Nevertheless, these gains in entropy or reductions in DBI often come at the cost of sacrificing dynamical properties, resulting in notably poor performance in terms of GMRQ scores and metastability compared to SPIB. Thus, one conclusion from the quantitative analysis presented in Table \ref{tab:large_dt_quantitative_comparison} is that SPIB generates state-of-the-art performance across four diverse quantitative scoring metrics measuring the quality of the generated kinetic model for all three protein folding systems.

In pursuit of a deeper understanding of the underlying dynamics with higher temporal and spatial resolution, we opt for a moderate $\Delta t$ to identify additional metastable states using SPIB. Specifically, to obtain about 10 metastable states through SPIB, we set $\Delta t = 10$ ns for Trp-cage and WW-domain, and $\Delta t = 20$ ns for HP35. All quantitative comparison results are consolidated in Table \ref{tab:moderate_dt_quantitative_comparison}. Similar to the case of large $\Delta t$, compared to tICA-PCCA+ and VAMPnet, SPIB demonstrates competitive or slightly superior performance in validation GMRQ scores and metastability across all three systems. More notably, SPIB excels in learning a greater number of well-populated states, leading to higher entropy scores on both the train and validation sets for all three proteins. In contrast, tICA-PCCA+ and VAMPnets tend to result in lower entropy, suggesting the capture of numerous sparsely populated states which is a consequence of an excessive emphasis on slow dynamics. Additionally, PCA-PCCA+ consistently exhibits the lowest DBI, indicating the presence of the most structurally distinct states.

\begin{figure}[htbp]
     \centering
     \begin{subfigure}[b]{0.22\textwidth}
         \caption{Trp-cage}
         \includegraphics[width=\textwidth]{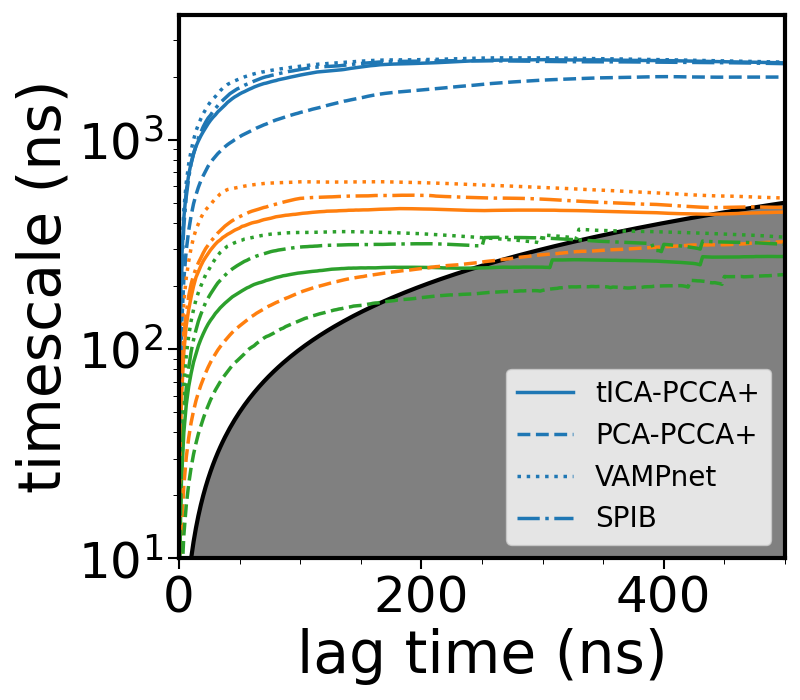}
     \end{subfigure}
     \hfill
     \begin{subfigure}[b]{0.22\textwidth}
         \caption{}
         \includegraphics[width=\textwidth]{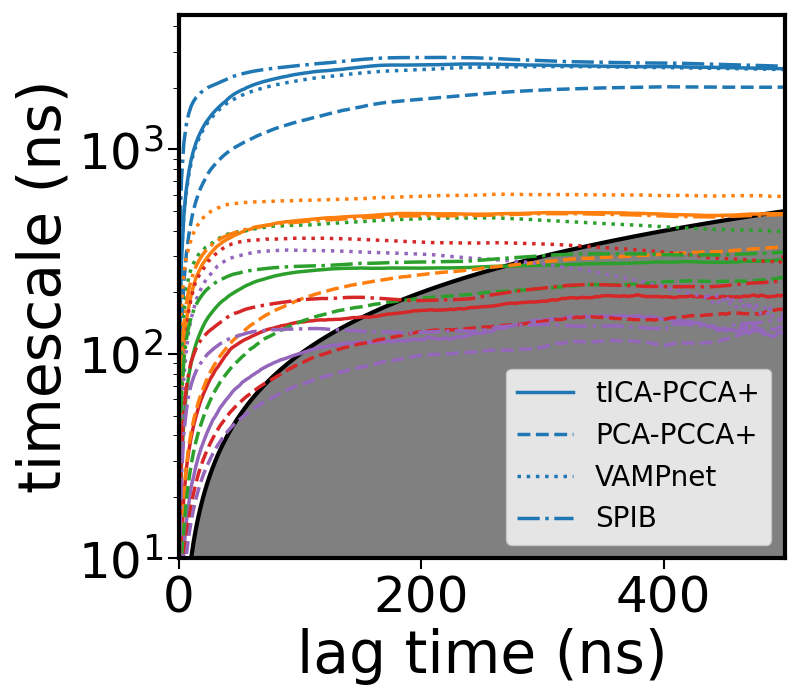}
     \end{subfigure}
     \hfill
     \begin{subfigure}[b]{0.22\textwidth}
         \caption{HP35}
         \includegraphics[width=\textwidth]{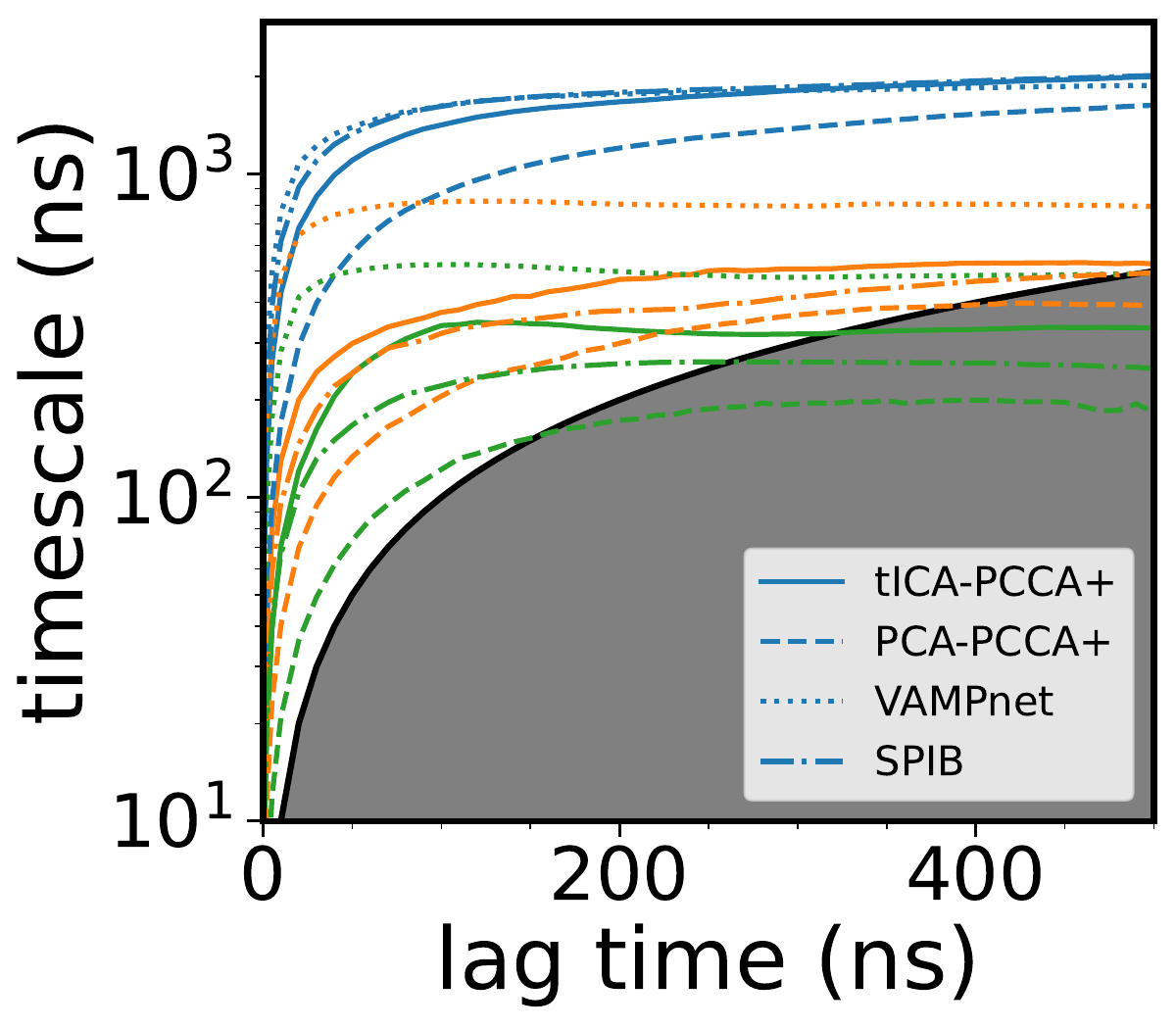}
     \end{subfigure}
     \hfill
     \begin{subfigure}[b]{0.22\textwidth}
         \caption{}
         \includegraphics[width=\textwidth]{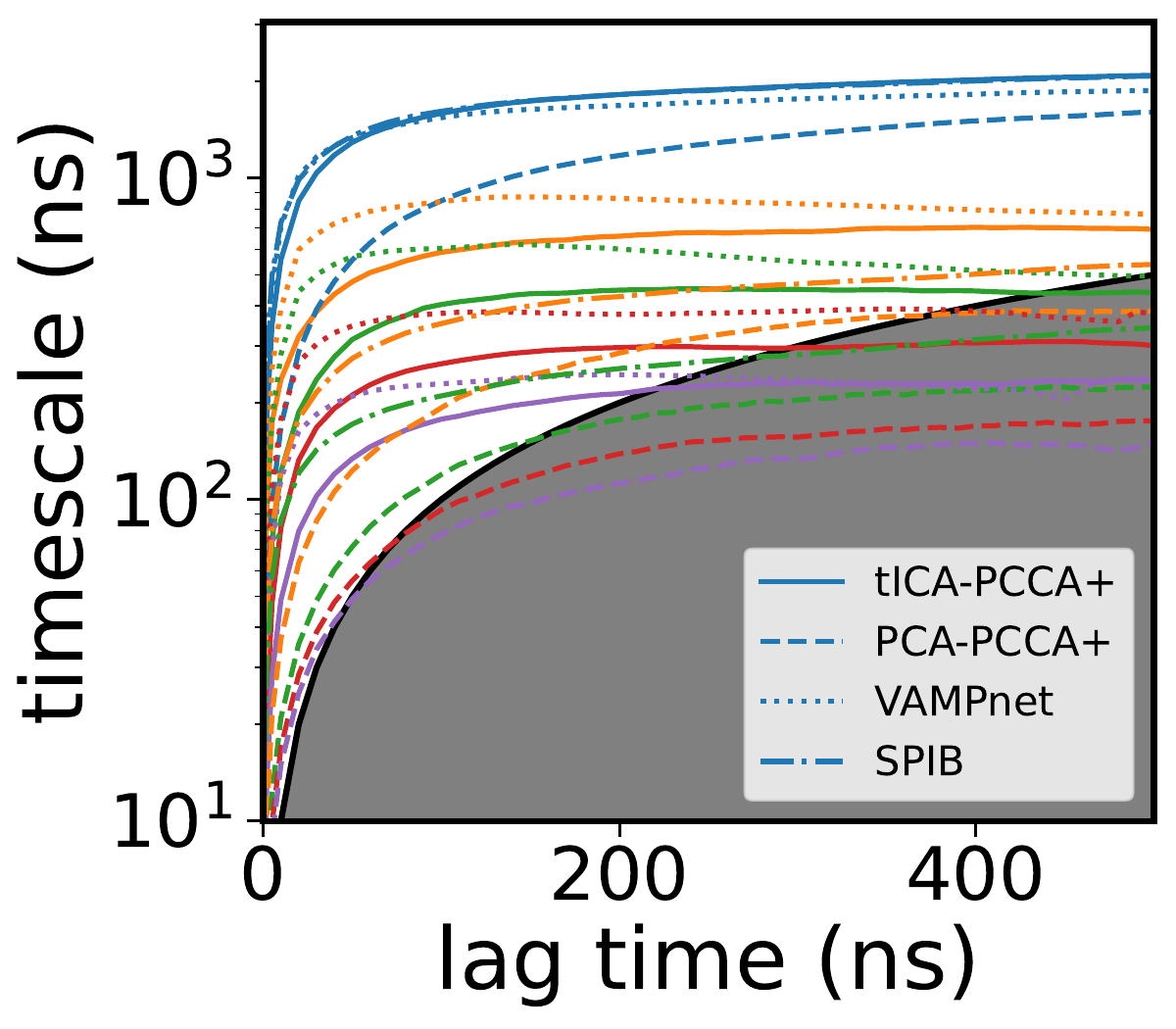}
     \end{subfigure}
     \hfill
     \begin{subfigure}[b]{0.22\textwidth}
         \caption{WW-domain}
         \includegraphics[width=\textwidth]{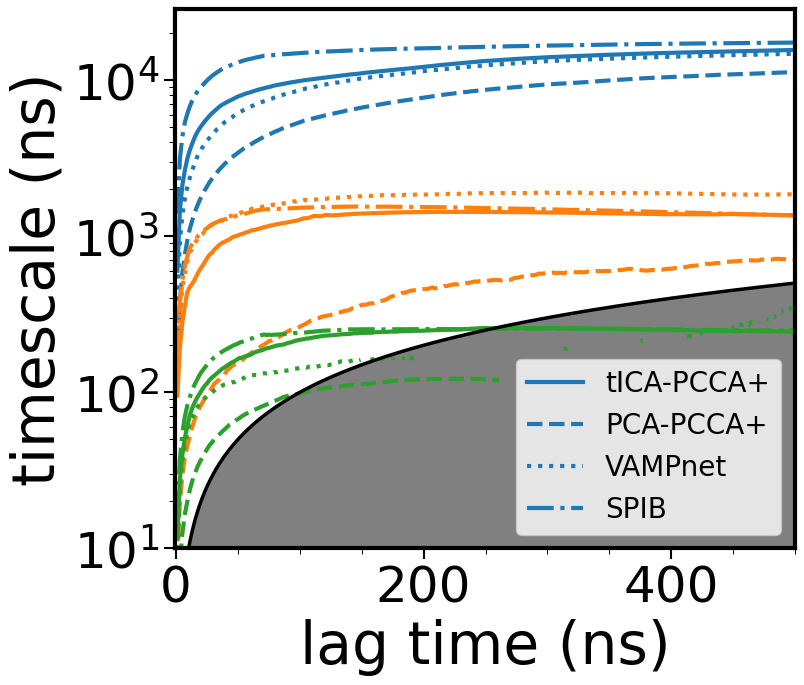}
     \end{subfigure}
     \hfill
     \begin{subfigure}[b]{0.22\textwidth}
         \caption{}
         \includegraphics[width=\textwidth]{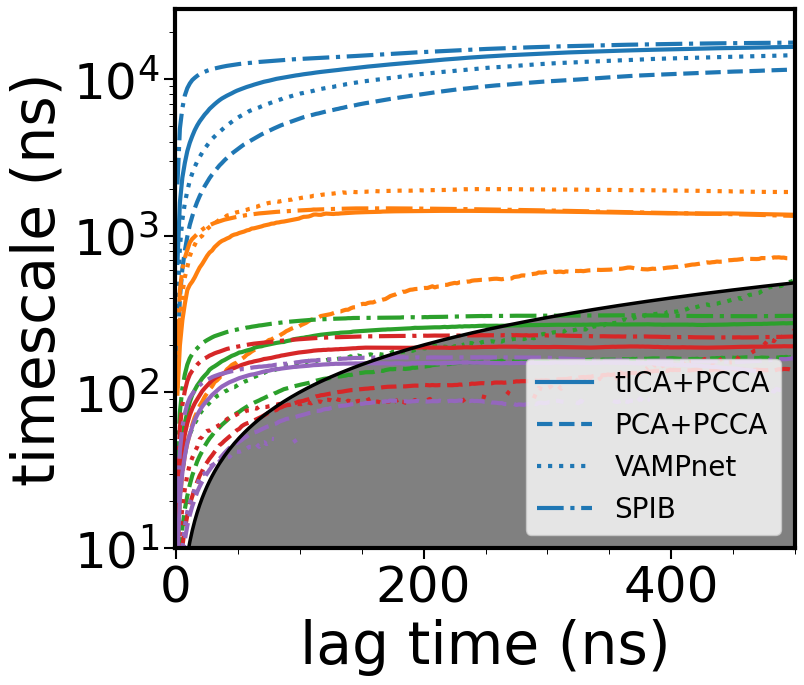}
     \end{subfigure}
        \caption{Implied timescales as a function of lag time for the MSMs of all systems. The left panels illustrate the results for 4-state MSMs in Trp-cage and 5-state MSMs in HP35 and WW-domain, while the right panels showcase the outcomes for 10-state MSMs. For clarity in presentation, only the mean values from 10 bootstrapping samples are plotted, with details available in the SI. The shaded gray area represents the region where timescales become equal to or smaller than the lag time and can no longer be resolved.}
        \label{fig:log_its_bootstrapping}
\end{figure}

To comprehensively compare the Markovian properties among the MSMs constructed by PCA-PCCA+, tICA-PCCA+, VAMPnets, and SPIB, we provide a detailed analysis in Fig. \ref{fig:log_its_bootstrapping}, which provides a close-up view of the ITS convergence as a function of lag time. The ITS are visualized using the mean value of 10 rounds of bootstrapping, with data randomly sampled with replacement. SPIB consistently achieves short Markovian lag times and large converged timescales especially for the slowest process, rendering its performance competitive with VAMPnet. However, in Trp-cage and HP35, VAMPnet outperforms SPIB for the other slow processes. It's worth noting that this superiority could be attributed to potential overfitting in VAMPnets, as SPIB consistently demonstrates comparable GMRQ scores in the validation set (Table. \ref{tab:large_dt_quantitative_comparison} and \ref{tab:moderate_dt_quantitative_comparison}). This suggests one potential advantage of SPIB: VAMP-based methods can have a higher likelihood of overfitting, particularly when employing a large number of states, as singular functions are estimated with inherent statistical uncertainty. In contrast, SPIB is trained in a self-consistent manner, which tends to be more robust and stable.

Across all three systems, SPIB outperforms tICA-PCCA+ and PCA-PCCA+ methods in the ITS, as the latter two fail to converge at the same lag times, and the converged timescales are slightly smaller. This observed trend is attributed to the capacity of SPIB to learn nonlinear transformations of input coordinates, providing enhanced resolution of slower processes, and the use of a continuous basis set for MSM macrostates construction. These enhanced capabilities yield a reduction of discretization errors compared to the Galerkin method when approximating the dynamical propagator. This results in higher converged timescales and the generation of state models with clearer time separations, leading to shorter relaxation times within states and, consequently, a reduced Markovian lag time.

\begin{table*}[htbp]
\caption{Quantitative comparison of different methods for large $\Delta t$ in terms of GMRQ (scoring based on the top 3 eigenvalues), metastability \revone{$Q$}, Shannon entropy, and DBI across three systems. \revtwo{The arrows indicate whether larger or smaller values are better for each metric.} The reported values represent the mean along with the standard error of the mean derived from 10-fold cross-validation results.}
\label{tab:large_dt_quantitative_comparison}
\centering
 \begin{tabular}{>{\centering\arraybackslash}p{2.02cm}| >{\centering\arraybackslash}p{1.8cm} >{\centering\arraybackslash}p{2.0cm} >{\centering\arraybackslash}p{1.9cm} >{\centering\arraybackslash}p{2.05cm} |>{\centering\arraybackslash}p{1.7cm} >{\centering\arraybackslash}p{1.7cm} >{\centering\arraybackslash}p{1.7cm} >{\centering\arraybackslash}p{1.7cm}} 
 \hline
 \multirow{2}{*}{\textbf{Model}} & \multicolumn{4}{c}{Trp-cage Train} & \multicolumn{4}{c}{Trp-cage Validation}\\
  & GMRQ $\uparrow$ & Q $\uparrow$ & H $\uparrow$ & DBI $\downarrow$ & GMRQ $\uparrow$ & Q $\uparrow$ & H $\uparrow$ & DBI $\downarrow$\\
 \hline
 \\[-1em]
    PCA-PCCA+ & $3.00\pm0.02$ & $0.749\pm0.004$ & $\mathbf{0.895\pm0.006}$ & $1.807\pm0.006$ & $2.72\pm0.08$ & $0.68\pm0.02$ & $\mathbf{0.86\pm0.04}$ & $1.85\pm0.03$\\
    tICA-PCCA+ & $3.42\pm0.01$ & $0.855\pm0.002$ & $0.837\pm0.005$ & $\mathbf{1.748\pm0.003}$ & $2.9\pm0.2$ & $\mathbf{0.76\pm0.03}$ & $0.79\pm0.05$ & $1.81\pm0.07$\\
    PCA-MPP & $2.46\pm0.03$ & $0.56\pm0.02$ & $0.71\pm0.02$ & $2.0\pm0.1$ & $2.17\pm0.09$ & $0.50\pm0.03$ & $0.68\pm0.05$ & $2.1\pm0.1$\\
    tICA-MPP & $2.79\pm0.08$ & $0.67\pm0.02$ & $0.51\pm0.01$ & $2.14\pm0.05$ & $2.06\pm0.09$ & $0.50\pm0.04$ & $0.47\pm0.05$ & $2.1\pm0.2$\\
    VAMPnet & $\mathbf{3.55\pm0.01}$ & $\mathbf{0.888\pm0.002}$ & $0.807\pm0.006$ & $1.822\pm0.003$ & $\mathbf{3.0\pm0.1}$ & $\mathbf{0.76\pm0.03}$ & $0.76\pm0.05$ & $1.82\pm0.04$\\
    SPIB & $3.51\pm0.01$ & $0.878\pm0.002$ & $0.797\pm0.006$ & $1.810\pm0.003$ & $\mathbf{3.0\pm0.1}$ & $0.75\pm0.03$ & $0.75\pm0.05$ & $\mathbf{1.76\pm0.02}$\\
 \hline
 \multirow{2}{*}{\textbf{Model}} & \multicolumn{4}{c}{HP35 Train} & \multicolumn{4}{c}{HP35 Validation}\\
  & GMRQ $\uparrow$ & Q $\uparrow$ & H $\uparrow$ & DBI $\downarrow$ & GMRQ $\uparrow$ & Q $\uparrow$ & H $\uparrow$ & DBI $\downarrow$\\
 \hline
 \\[-1em]
    PCA-PCCA+ & $3.00\pm0.02$ & $0.73\pm0.01$ & $0.65\pm0.08$ & $3.5\pm0.2$ & $2.3\pm0.2$ & $0.62\pm0.08$ & $0.8\pm0.1$ & $3.5\pm0.2$\\
    tICA-PCCA+ & $3.57\pm0.02$ & $0.80\pm0.02$ & $0.92\pm0.05$ & $5.3\pm0.3$ & $3.0\pm0.1$ & $\mathbf{0.69\pm0.04}$ & $0.87\pm0.06$ & $4.6\pm0.6$\\
    PCA-MPP & $3.12\pm0.02$ & $0.64\pm0.03$ & $0.91\pm0.08$ & $\mathbf{2.5\pm0.1}$ & $2.5\pm0.2$ & $0.28\pm0.02$ & $0.89\pm0.08$ & $\mathbf{2.69\pm0.08}$\\
    tICA-MPP & $3.52\pm0.02$ & $0.76\pm0.02$ & $1.21\pm0.05$ & $3.6\pm0.3$ & $\mathbf{3.1\pm0.1}$ & $\mathbf{0.69\pm0.03}$ & $1.18\pm0.04$ & $3.1\pm0.3$\\
    VAMPnet & $\mathbf{3.65\pm0.04}$ & $\mathbf{0.83\pm0.03}$ & $0.8\pm0.2$ & $3.5\pm0.2$ & $2.9\pm0.1$ & $0.65\pm0.02$ & $0.77\pm0.07$ & $3.1\pm0.2$\\
    SPIB & $3.51\pm0.02$ & $\mathbf{0.83\pm0.01}$ & $\mathbf{1.26\pm0.01}$ & $3.74\pm0.04$ & $3.0\pm0.1$ & $0.67\pm0.03$ & $\mathbf{1.19\pm0.03}$ & $3.4\pm0.1$\\
 \hline
 \multirow{2}{*}{\textbf{Model}} & \multicolumn{4}{c}{WW-domain Train} & \multicolumn{4}{c}{WW-domain Validation}\\
  & GMRQ $\uparrow$ & Q $\uparrow$ & H $\uparrow$ & DBI $\downarrow$ & GMRQ $\uparrow$ & Q $\uparrow$ & H $\uparrow$ & DBI $\downarrow$\\
 \hline
 \\[-1em]
    PCA-PCCA+ & $3.02\pm0.01$ & $0.652\pm0.006$ & $0.587\pm0.006$ & $\mathbf{1.83\pm0.03}$ & $2.4\pm0.1$ & $0.49\pm0.03$ & $0.57\pm0.04$ & $\mathbf{1.76\pm0.05}$\\
    tICA-PCCA+ & $3.48\pm0.01$ & $0.788\pm0.006$ & $0.591\pm0.006$ & $2.27\pm0.02$ & $2.7\pm0.2$ & $0.58\pm0.04$ & $0.56\pm0.04$ & $2.19\pm0.09$\\
    PCA-MPP & $2.35\pm0.06$ & $0.46\pm0.03$ & $\mathbf{0.9\pm0.1}$ & $8.5\pm0.8$ & $2.05\pm0.07$ & $0.41\pm0.03$ & $\mathbf{0.91\pm0.09}$ & $7.9\pm0.6$\\
    tICA-MPP & $3.39\pm0.05$ & $0.79\pm0.02$ & $0.592\pm0.006$ & $1.99\pm0.05$ & $2.6\pm0.2$ & $0.60\pm0.05$ & $0.56\pm0.03$ & $1.8\pm0.1$\\
    VAMPnet & $\mathbf{3.64\pm0.01}$ & $\mathbf{0.841\pm0.003}$ & $0.592\pm0.003$ & $2.22\pm0.02$ & $\mathbf{2.8\pm0.2}$ & $0.58\pm0.04$ & $0.56\pm0.04$ & $2.0\pm0.1$\\
    SPIB & $3.59\pm0.01$ & $0.823\pm0.003$ & $0.627\pm0.003$ & $2.29\pm0.07$ & $\mathbf{2.8\pm0.2}$ & $\mathbf{0.67\pm0.03}$ & $0.61\pm0.05$ & $2.28\pm0.09$\\
 \hline
\end{tabular}
\end{table*}

\begin{table*}[htbp]
\caption{Quantitative comparison of different methods for moderate $\Delta t$ in terms of GMRQ (scoring based on the top 5 eigenvalues), metastability \revone{$Q$}, Shannon entropy, and DBI across three systems. The reported values represent the mean along with the standard error of the mean derived from 10-fold cross-validation results.}
\label{tab:moderate_dt_quantitative_comparison}
\centering
 \begin{tabular}{>{\centering\arraybackslash}p{2.02cm}| >{\centering\arraybackslash}p{1.8cm} >{\centering\arraybackslash}p{2.0cm} >{\centering\arraybackslash}p{2.0cm} >{\centering\arraybackslash}p{1.8cm} |>{\centering\arraybackslash}p{1.7cm} >{\centering\arraybackslash}p{1.7cm} >{\centering\arraybackslash}p{1.7cm} >{\centering\arraybackslash}p{1.7cm}} 
 \hline
 \multirow{2}{*}{\textbf{Model}} & \multicolumn{4}{c}{Trp-cage Train} & \multicolumn{4}{c}{Trp-cage Validation}\\
  & GMRQ $\uparrow$ & Q $\uparrow$ & H $\uparrow$ & DBI $\downarrow$ & GMRQ $\uparrow$ & Q $\uparrow$ & H $\uparrow$ & DBI $\downarrow$\\
 \hline
 \\[-1em]
    PCA-PCCA+ & $3.42\pm0.02$ & $0.401\pm0.002$ & $1.33\pm0.02$ & $\mathbf{2.10\pm0.05}$ & $2.9\pm0.1$ & $0.31\pm0.01$ & $\mathbf{1.27\pm0.04}$ & $\mathbf{2.17\pm0.06}$\\
    tICA-PCCA+ & $4.22\pm0.02$ & $0.506\pm0.005$ & $1.00\pm0.03$ & $2.55\pm0.09$ & $3.1\pm0.1$ & $0.32\pm0.02$ & $0.92\pm0.05$ & $3.0\pm0.1$\\
    PCA-MPP & $2.96\pm0.03$ & $0.323\pm0.008$ & $1.16\pm0.04$ & $2.3\pm0.1$ & $2.39\pm0.08$ & $0.25\pm0.01$ & $1.12\pm0.07$ & $2.3\pm0.1$\\
    tICA-MPP & $3.52\pm0.07$ & $0.39\pm0.01$ & $0.66\pm0.05$ & $4.3\pm0.4$ & $2.3\pm0.1$ & $0.26\pm0.01$ & $0.61\pm0.07$ & $4.5\pm0.3$\\
    VAMPnet & $\mathbf{5.05\pm0.02}$ & $\mathbf{0.735\pm0.006}$ & $1.00\pm0.01$ & $3.0\pm0.3$ & $3.0\pm0.1$ & $0.37\pm0.05$ & $0.86\pm0.06$ & $2.5\pm0.2$\\
    SPIB & $4.48\pm0.03$ & $0.538\pm0.009$ & $\mathbf{1.35\pm0.05}$ & $3.03\pm0.08$ & $\mathbf{3.5\pm0.1}$ & $\mathbf{0.40\pm0.02}$ & $1.25\pm0.08$ & $2.8\pm0.1$\\
 \hline
 \multirow{2}{*}{\textbf{Model}} & \multicolumn{4}{c}{HP35 Train} & \multicolumn{4}{c}{HP35 Validation}\\
  & GMRQ $\uparrow$ & Q $\uparrow$ & H $\uparrow$ & DBI $\downarrow$ & GMRQ $\uparrow$ & Q $\uparrow$ & H $\uparrow$ & DBI $\downarrow$\\
 \hline
 \\[-1em]
    PCA-PCCA+ & $3.27\pm0.08$ & $0.38\pm0.01$ & $\mathbf{1.53\pm0.02}$ & $\mathbf{2.8\pm0.1}$ & $2.6\pm0.2$ & $0.29\pm0.02$ & $\mathbf{1.50\pm0.05}$ & $\mathbf{2.91\pm0.08}$\\
    tICA-PCCA+ & $4.816\pm0.008$ & $0.619\pm0.004$ & $1.45\pm0.02$ & $5.2\pm0.2$ & $4.0\pm0.1$ & $\mathbf{0.49\pm0.02}$ & $1.40\pm0.05$ & $4.6\pm0.2$\\
    PCA-MPP & $3.83\pm0.02$ & $0.34\pm0.02$ & $1.2\pm0.1$ & $3.4\pm0.1$ & $2.8\pm0.2$ & $0.22\pm0.02$ & $1.1\pm0.1$ & $3.5\pm0.1$\\
    tICA-MPP & $4.47\pm0.06$ & $0.50\pm0.03$ & $1.27\pm0.03$ & $5.8\pm0.8$ & $\mathbf{4.1\pm0.8}$ & $0.47\pm0.03$ & $1.24\pm0.04$ & $3.9\pm0.4$\\
    VAMPnet & $\mathbf{5.21\pm0.08}$ & $0.61\pm0.02$ & $1.41\pm0.04$ & $4.1\pm0.3$ & $3.4\pm0.3$ & $0.37\pm0.02$ & $1.37\pm0.05$ & $4.3\pm0.2$\\
    SPIB & $5.02\pm0.04$ & $\mathbf{0.65\pm0.02}$ & $1.511\pm0.009$ & $4.7\pm0.1$ & $3.8\pm0.2$ & $0.46\pm0.03$ & $1.44\pm0.05$ & $3.8\pm0.1$\\
 \hline
 \multirow{2}{*}{\textbf{Model}} & \multicolumn{4}{c}{WW-domain Train} & \multicolumn{4}{c}{WW-domain Validation}\\
  & GMRQ $\uparrow$ & Q $\uparrow$ & H $\uparrow$ & DBI $\downarrow$ & GMRQ $\uparrow$ & Q $\uparrow$ & H $\uparrow$ & DBI $\downarrow$\\
 \hline
 \\[-1em]
    PCA-PCCA+ & $3.72\pm0.01$ & $0.44\pm0.02$ & $0.67\pm0.01$ & $\mathbf{2.13\pm0.04}$ & $2.9\pm0.1$ & $0.32\pm0.02$ & $0.65\pm0.04$ & $\mathbf{2.11\pm0.06}$\\
    tICA-PCCA+ & $4.65\pm0.01$ & $0.622\pm0.004$ & $0.773\pm0.005$ & $4.72\pm0.03$ & $\mathbf{3.7\pm0.2}$ & $0.43\pm0.02$ & $0.75\pm0.06$ & $4.30\pm0.08$\\
    PCA-MPP & $2.37\pm0.05$ & $0.221\pm0.007$ & $\mathbf{1.2\pm0.1}$ & $7.9\pm0.5$ & $2.06\pm0.07$ & $0.19\pm0.01$ & $\mathbf{1.1\pm0.1}$ & $7.7\pm0.5$\\
    tICA-MPP & $3.51\pm0.01$ & $0.54\pm0.01$ & $0.67\pm0.01$ & $5.0\pm0.4$ & $2.9\pm0.2$ & $0.39\pm0.02$ & $0.65\pm0.04$ & $4.2\pm0.4$\\
    VAMPnet & $\mathbf{4.91\pm0.02}$ & $\mathbf{0.669\pm0.006}$ & $0.75\pm0.01$ & $4.18\pm0.07$ & $3.4\pm0.2$ & $\mathbf{0.44\pm0.03}$ & $0.72\pm0.05$ & $3.7\pm 0.1$\\
    SPIB & $4.75\pm0.02$ & $0.57\pm0.01$ & $0.95\pm0.02$ & $6.5\pm0.4$ & $3.6\pm0.1$ & $0.42\pm0.02$ & $0.92\pm0.05$ & $5.4\pm0.3$\\
 \hline
\end{tabular}
\end{table*}

\subsection{Elucidating biophysical mechanisms}

Having established quantitatively the SPIB method as a viable state-of-the-art method for the construction of kinetic models whose performance is on par with other established MSM-construction methodologies, we now analyze the protein folding dynamics based on the macrostates constructed and validated from SPIB in the previous subsection. This qualitative analysis provides insights regarding the macrostates predicted by SPIB and compares these states with those obtained using other methods. 

We show how SPIB is able to learn the important folded, misfolded, and unfolded states of each mini-protein at both the coarse (large $\Delta t$) and fine (moderate $\Delta t$) levels, while other state-of-the-art methods sometimes erroneously lump these conformationally distinct states together. We show the learned SPIB latent spaces and metastable states at both coarse and fine resolutions as well as the flux network for the fine SPIB models. This transition from coarse to fine SPIB models demonstrates a refined understanding of folding and unfolding mechanisms. Finally, we provide a brief summary comparing SPIB states with those predicted by other methods, with further details available in the SI.

\subsubsection{Trp-Cage}

\begin{figure}[t!]
    \centering
    \includegraphics[width=0.45\textwidth]{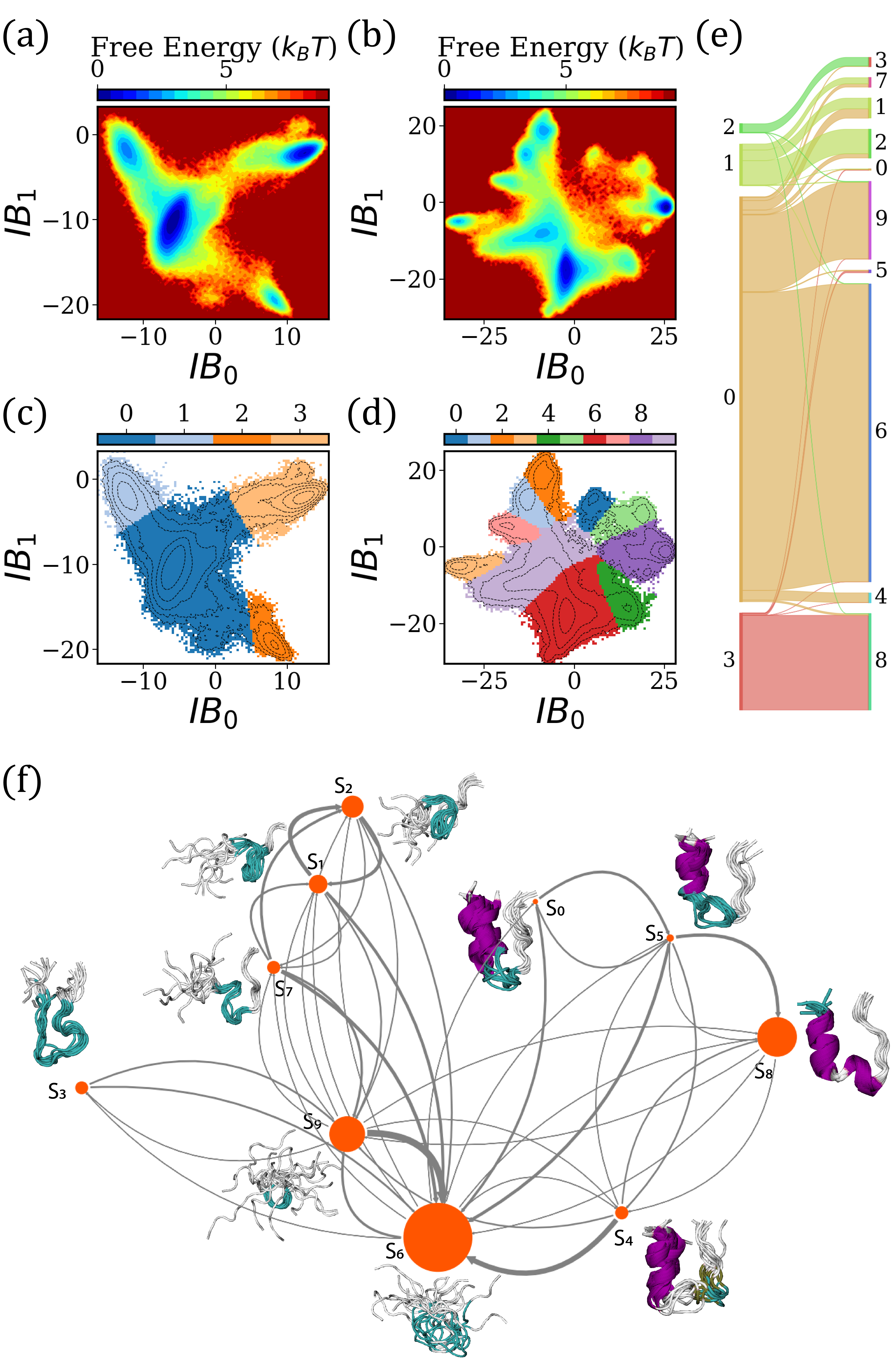}
    \caption{Qualitative description of the MSM analysis for Trp-cage protein. (a) and (b) give the free-energy surfaces in the two-dimensional SPIB latent space\revtwo{, denoted by $\text{IB}_0$ and $\text{IB}_1$,} for large and moderate $\Delta t$, respectively. (c) and (d) give the metastable states learned by SPIB in the case of large and moderate $\Delta t$, respectively. (e) The Sankey plot illustrates the corresponding relations between states learned by SPIB using large (left) and moderate (right) $\Delta t$. (f) The MSM constructed based on states identified by SPIB, trained with a moderate $\Delta t$, is visualized using a flux network. The node size is proportional to the stationary population of the states, and the arrow width is scaled according to jump probabilities. Additionally, ten randomly selected conformations from each state are overlapped and displayed adjacent to the corresponding node.}
    \label{fig:trpcage_qualitative_comparison}
\end{figure}

The Trp-cage protein, comprising 20 residues, stands as one of the well-known examples of small-sized folding proteins. It folds to a native state characterized by an N-terminal $\alpha$-helix, followed by a short $3_{10}$-helix, a C-terminal polyproline II region, and a hydrophobic core stabilized by interactions of the Trp6 side chain with Pro12, Pro18, and Pro19, as illustrated in Fig. \ref{fig:systems}. Fig. \ref{fig:trpcage_qualitative_comparison} illustrates how SPIB adeptly identifies metastable states in the system across different levels of coarse-graining. When employing a large lag time $\Delta t$, SPIB discerns and represents the system with 4 states, as illustrated in Fig. \ref{fig:trpcage_qualitative_comparison}(a,c). Meanwhile, with a moderate lag time, SPIB captures a more detailed picture by learning 10 states, as showcased in Fig. \ref{fig:trpcage_qualitative_comparison}(b, d). The relationship between the 4-state and 10-state models is effectively elucidated through a Sankey plot presented in Fig. \ref{fig:trpcage_qualitative_comparison}(e). This Sankey plot visually illustrates the relationship between two sets of states by mapping one set of states to another set of states, revealing the hierarchical arrangement of metastable states.\cite{Nagel2023a}

Under large $\Delta t$, SPIB yields a minimal MSM for analysis. The salient features of the landscape include a folded state represented by state 3, intricately linked to the molten globule state (state 0) through a narrow bottleneck. State 0, in turn, connects to the unfolded state (state 1), characterized by multiple turns in the structure, via a small energy barrier, and to a hairpin state (state 2) through a significant energy barrier.

As the lag time $\Delta t$ is reduced, the initial 4-state model undergoes further refinement, evolving into a more detailed 10-state model, as depicted in Fig. \ref{fig:trpcage_qualitative_comparison}(e). The metastable conformational ensembles and associated transitions for the 10 macrostates are visually represented in Fig. \ref{fig:trpcage_qualitative_comparison}(f). Notably, S$_8$ corresponds to the folded state, while S$_0$ and S${_5}$ represent intermediate states bridging the folded and unfolded states, featuring a partially folded N-terminus. Additionally, S$_9$ embodies a crossed conformation with a minor central hairpin, and S$_6$ comprises a blend of molten globule structures and an extended conformation. S$_4$ signifies a partially unfolded state. Furthermore, S$_1$ and S$_2$ adopt a braided hairpin-like structure, S$_7$ manifests as a partially compact configuration with multiple turns, and S$_3$ exhibits a distinct hairpin conformation. These identified states resemble those reported in literature, confirming their consistency and relevance.\cite{Sidky2019} A clear correspondence emerges between the SPIB-learned latent space, as depicted in Fig. \ref{fig:trpcage_qualitative_comparison}(b,d), and the constructed MSM network shown in Fig. \ref{fig:trpcage_qualitative_comparison}(f). This observation suggests that SPIB actually learns a continuous embedding of the MD conformations, serving as an information bottleneck that maximally preserves information about state-to-state transitions. 

A comprehensive examination of the consequences of utilizing alternative methods for macrostate construction is provided in the SI. Here, we highlight the key findings. For the 4-state model, SPIB aligns closely with VAMPnets, while tICA-PCCA+ and PCA-PCCA+ identify states but struggle with precise boundaries. In contrast, PCA-MPP and tICA-MPP struggle to correctly identify or distinguish the unfolded ensemble of states, often missing one or two important unfolded states. For the 10-state model, SPIB excels in capturing a more refined model with numerous well-populated macrostates. This contrasts with other methods, which struggle to further subdivide highly populated states, leading to the emergence of numerous lowly populated states. While applying MPP and PCCA+ on PCA appears to alleviate this issue, their overall performance in dynamical metrics is generally suboptimal.

\subsubsection{HP35}

\begin{figure}[t!]
    \centering
    \includegraphics[width=0.45\textwidth]{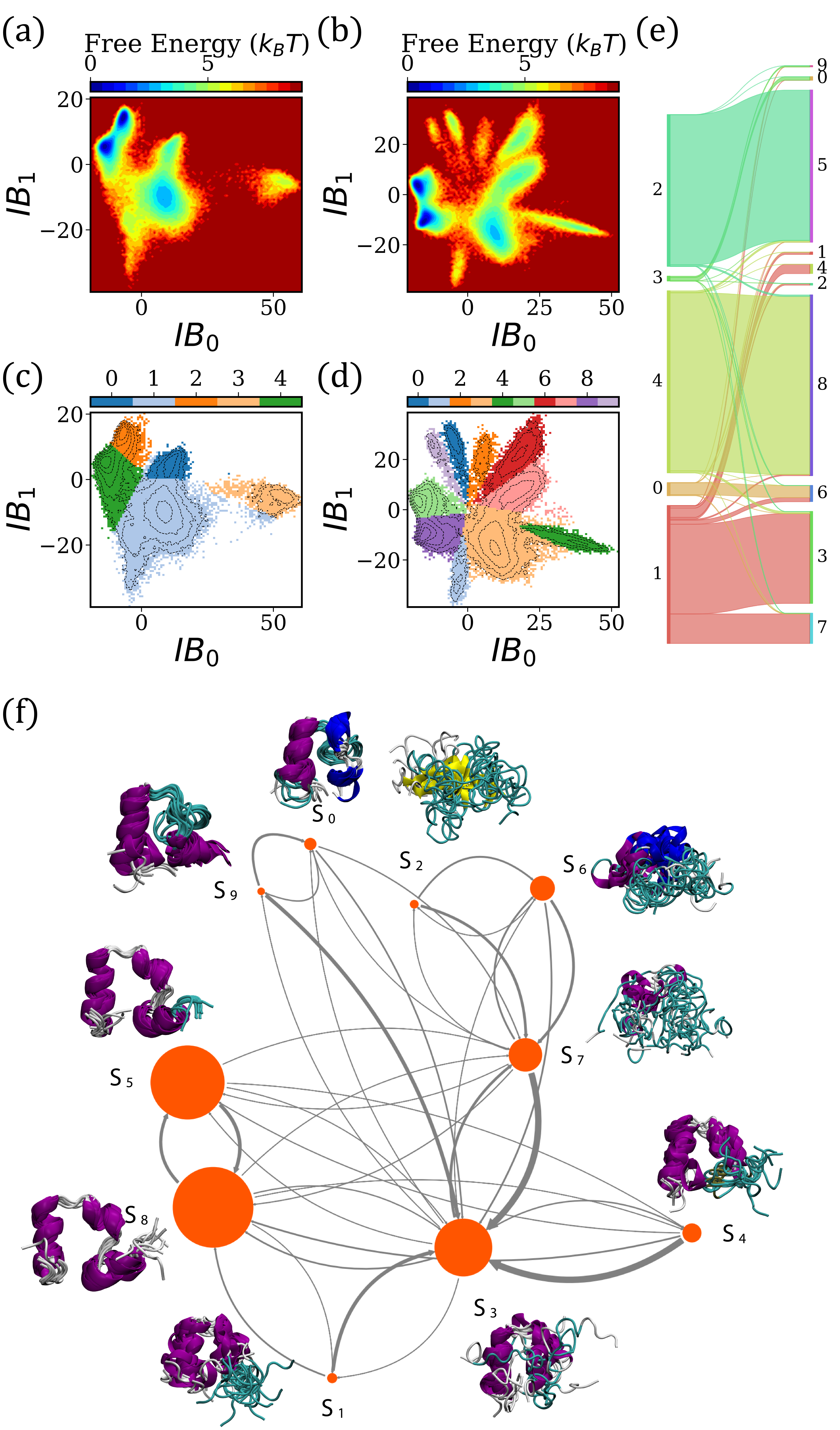}
    \caption{Qualitative description of the MSM analysis for HP35 protein. (a) and (b) give the free-energy surfaces in the two-dimensional SPIB latent space for large (100 ns) and moderate (20 ns) $\Delta t$, respectively. (c) and (d) give the metastable states learned by SPIB in the case of large and moderate $\Delta t$, respectively. (e) The Sankey plot illustrates the corresponding relations between states learned by SPIB using large (left) and moderate (right) $\Delta t$. (f) The MSM constructed based on states identified by SPIB, trained with a moderate $\Delta t$, is visualized using a flux network. The node size is proportional to the stationary population of the states, and the arrow width is scaled according to jump probabilities. Additionally, ten randomly selected conformations from each state are overlapped and displayed adjacent to the corresponding node, with the secondary structure for each frame templated on a single, randomly selected frame from all ten.}
    \label{fig:hp35_qualitative_comparison}
\end{figure}

Fig. \ref{fig:hp35_qualitative_comparison} shows the analogous qualitative results for HP35. Fig. \ref{fig:hp35_qualitative_comparison}(a,b) show the free-energy surfaces in the SPIB latent space for a large (100 ns) and moderate (20 ns) values of $\Delta t$, indicating the kinetic model at a coarse and fine resolution, respectively. Fig. \ref{fig:hp35_qualitative_comparison}(c,d) shows the learned SPIB metastable partitioning for these free-energy surfaces; as expected, using the smaller value of the lag time in the method allows for learning a larger number of metastable states. Fig. \ref{fig:hp35_qualitative_comparison}(e) shows via a Sankey plot the topological mapping of the states learned at large $\Delta t$ to the states learned at moderate $\Delta t$. The topological changes mostly involve a finer partitioning of the unfolded state (state 1 in the 5-state model) learned at large $\Delta t$ into a series of substates while the folded states are left mostly intact.

The connectivity network of the discrete state model learned by SPIB at moderate $\Delta t$ is given in Fig. \ref{fig:hp35_qualitative_comparison}(f), where state-to-state transition network is diagrammed schematically. This network shows that the two folded states S$_5$ and S$_8$ are the most populated, and they differ by changes in the $\phi_3$ dihedral angle (Fig. \ref{fig:systems}, middle), which has been shown previously to be a sufficient coordinate to distinguish the folded states of HP35\cite{Nagel2023a}, which are S$_5$ ($\phi_3 > 0$ rad) and S$_8$ ($\phi_3 < 0$ rad) in our 10 state analysis and states 2 and 4  in the 5 state analysis.

The other 8 states in the 10 state model correspond to states with various degrees of unfolding or misfolded states. Unfolded state 1 in the 5 state model is decomposed into states S$_1$, S$_2$, S$_3$, S$_4$, and S$_7$ in the 10 state model. All these unfolded states contain some degree of secondary structure, but differ in the fraction and location of the structure. Specifically, state S$_1$ presents with alpha helix 1 ($\alpha_1$) folded, $\alpha_2$ unfolded, and $\alpha_3$ misfolded, S$_2$ with $\alpha_1$ unfolded and the other two helices misfolded, S$_3$ and S$_4$ with $\alpha_1$ unfolded and the other two helices folded, and S$_7$ presenting as a completely unfolded state.

State S$_3$ has $\alpha_1$ unfolded and other two helices $\alpha_2$ and $\alpha_3$ folded and serves as a hub on the folding trajectory, demonstrating that the 10-state SPIB model predicts that $\alpha_1$ folds last. A flux analysis using transition path theory starting from the unfolded state S$_2$ and ending in folded state S$_8$ tentatively suggests that $\alpha_3$ folds first, followed by $\alpha_2$, then $\alpha_1$. States S$_0$ and S$_9$, where helix $\alpha_1$ folds before helix $\alpha_2$ lie off the major folding flux and serve as misfolded trap states in our analysis.

Generally, S$_7$ appears to be a sink for both the misfolded state S$_1$ and the other unfolded states. However, overall, the qualitative analysis of the 10 state model indicates that there is not a dominant folding-unfolding pathway for the HP35 protein; it instead indicates that, even for this simple protein, the folding processes are multifaceted without a single dominant route. 

A comparison of the SPIB metastable states at both the 5 and 10 state level of resolution and those discovered by the PCCA+, MPP, and VAMPnet approaches are given in the SI. There we observe that only SPIB and tICA-MPP are able to distinguish clearly and completely the two folded states discriminated by the $\phi_3$ dihedral while all other methodologies tested lump those two states into one metastable, folded state at the coarser 5 state level. When the clustering is performed at the finer 10 state level, all methods are able to distinguish the two folded states differing in $\phi_3$ angle, although the VAMPnet approach only resolves the two folded states for 40\% of the models built. This result is consistent with VAMPnet's occasional failure in predicting the third-slowest timescale in an alanine dipeptide system\cite{Mardt2018}.

\subsubsection{WW domain}

\begin{figure}[t!]
    \centering
    \includegraphics[width=0.45\textwidth]{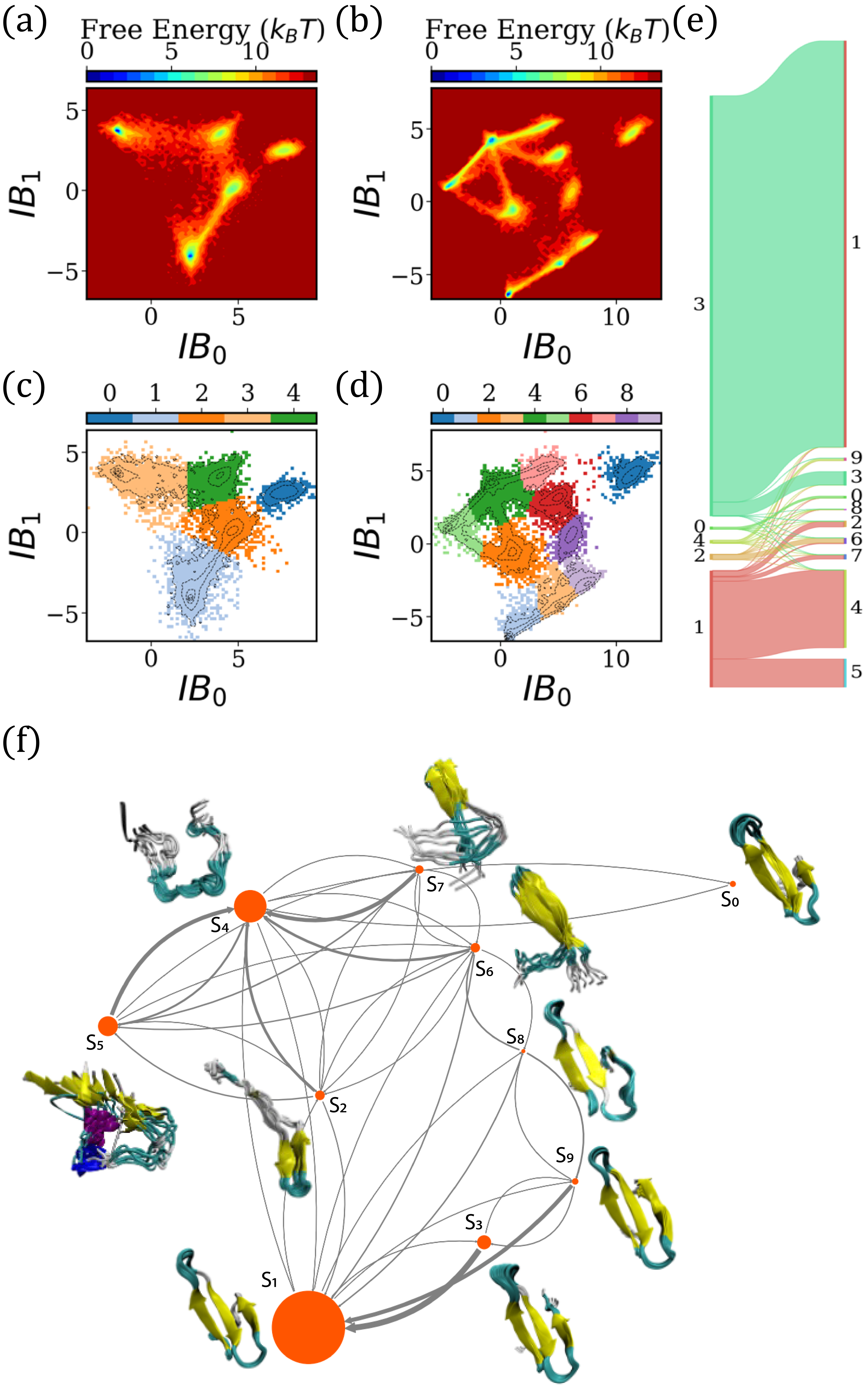}
    \caption{Qualitative description of the MSM analysis for WW-domain protein. (a) and (b) give the free-energy surfaces in the two-dimensional SPIB latent space for large and moderate $\Delta t$, respectively. (c) and (d) give the metastable states learned by SPIB in the case of large and moderate $\Delta t$, respectively. (e) The Sankey plot illustrates the corresponding relations between states learned by SPIB using large (left) and moderate (right) $\Delta t$. (f) The MSM constructed based on states identified by SPIB, trained with a moderate $\Delta t$, is visualized using a flux network. The node size is proportional to the stationary population of the states, and the arrow width is scaled according to jump probabilities. Additionally, ten randomly selected conformations from each state are overlapped and displayed adjacent to the corresponding node.}
    \label{fig:wwdomain_qualitative_comparison}
\end{figure}

The WW-domain protein consists of 35 residues which could form a three-stranded beta-sheet, with residues 8-23 forming hairpin 1 and residues 17-30 forming hairpin 2 (as shown in Fig. \ref{fig:systems}). Recent investigations, employing both experimental techniques and MD simulations, have explored the folding mechanism of the WW-domain. Two distinct folding mechanisms have been elucidated, differing in the folding order of hairpin 1 and hairpin 2. Approximately $70\%$ of the WW-domain protein folding process involves the sequential folding of hairpin 1 followed by hairpin 2, while the remaining $30\%$ undergoes folding in the opposite order\cite{qiu2023efficient, noe2009constructing, weikl2008transition, a2012dominant}. In this study, adopting SPIB for MSMs analysis, we obtained results qualitatively consistent with previous studies. 

Using SPIB to construct MSMs for the WW-domain protein folding system, Fig. \ref{fig:wwdomain_qualitative_comparison} illustrates the outcomes produced by two SPIB models trained at $\Delta t=$ 70 ns and 10 ns, respectively. Setting $\Delta t$ to a large value resulted in a highly coarse-grained model consisting of 5 states. The latent space of the trained SPIB and the distributions of different states are visualized in Fig. \ref{fig:wwdomain_qualitative_comparison}(a) and (c). Using a much smaller $\Delta t$ of 10 ns, a model with higher resolution (i.e., 10 states) could be obtained, as depicted in Fig. \ref{fig:wwdomain_qualitative_comparison}(b) and (d). The representative conformations for the 10 states and the transition relationships between them are elucidated by the network flux plot in Fig. \ref{fig:wwdomain_qualitative_comparison}(e). The corresponding relations between the 5 states and the 10 states are illustrated in Fig.\ref{fig:wwdomain_qualitative_comparison} (f).

Clearly, both the 5-states and 10-states MSMs introduced by SPIB successfully identified the folded, unfolded, misfolded, and partially folded states.\cite{lane2011markov} In the 5-states MSM, state 0 corresponds to the mis-folded state, while states 1 and 3 correspond to the unfolded and folded states, respectively. State 2 represents the hairpin 1 partially folded state, while state 4 is intricately devised, encompassing both hairpin 1 and hairpin 2 partially folded conformations. The 5-states model effectively connects the unfolded and folding states through partially folded states and separates the misfolded state. However, due to the limited model resolution, the distinctions between the different folding mechanisms are not very evident.

In the 10-state MSM with higher resolution, the two folding mechanisms are clearly identified. States S$_4$ and S$_5$ represent the unfolded states, states S$_1$ and S$_3$ correspond to the folded states, and S$_0$ is identified as misfolded. Additionally, two distinct connections between unfolded and folded states emerge, where states 7-6-8-9 represent the hairpin 1 - hairpin 2 folding sequence, and states 2 show the hairpin 2 - hairpin 1 folding sequence. The widths of flux arrows also support the dominance of the hairpin 1 - hairpin 2 folding sequence. Interestingly, the folded and unfolded states are further separated into states with different interactions between the terminal parts of the protein. This model significantly enhances our comprehension of the folding process of the WW-domain protein.

More detailed comparisons between the states identified by SPIB and other methods can be found in the SI. In general, for the 5-states model, SPIB results are consistent with models constructed using tICA-PCCA+, tICA-MPP, and VAMPnets. Given the existence of multiple states with relatively low populations (i.e.,  $<1\%$) for the WW-domain, distinguishing the partially folded states proves challenging for PCA-based methods. In the 10-state model, while PCA-based methods still struggle to separate the kinetically stable states, VAMPnets face difficulties distinguishing differences within partial-folded and unfolded conformations. Results from the tICA-PCCA+ method are mostly consistent with SPIB results. Overall, SPIB demonstrates robust performance and effectively distinguished various meaningful metastable states, underscoring its significant potential for constructing MSMs.

\section{Discussion}

In this work, we employ PCA or tICA to identify a subspace within the input features and utilize k-means clustering to classify the conformations as the initial state labels for SPIB. For a large lag time $\Delta t$, we observe robustness in the results with respect to the choice of the initial label schemes. Both tICA+kmeans and PCA+kmeans initial state labels yield the same states. However, with a moderate lag time, these two schemes may result in different intermediate states and consequently, different outcomes. As a general recommendation, we suggest using tICA+kmeans for improved capture of slow dynamics, leading to better GMRQ and metastability. On the other hand, PCA+kmeans is preferred for learning more structurally distinct states, enhancing the DBI. As we aim to achieve superior dynamical properties, all results presented in the main text are obtained using tICA+kmeans to generate initial state labels. Detailed results and further discussion can be found in the SI.

In terms of SPIB hyperparameters, our observations indicate robust performance across a broad range of hyperparameter values (refer to results in the SI). However, for achieving an optimal MSMs performance, we recommend employing the GMRQ score and cross-validation for hyperparameter tuning. The GMRQ measures the ability of the model to capture the slowest dynamics, while the cross-validation procedure allows for a robust quantification of the model's performance on new data. Our recommended tuning process begins with fixing a large lag time $\Delta t$ to derive a small number of macrostates while subsequently tuning the other hyperparameters in a sequential manner. Once the optimization of all other hyperparameters is complete, the next step involves scanning the lag time $\Delta t$ for SPIB. This step will generate varying numbers of converged states for different $\Delta t$. To finally determine the lag time $\Delta t$, one can either select the plateau (if there is a clear timescale separation) or specify $\Delta t$ based on the time resolution that we care about in a dynamical system. Alternatively, one may choose the lag time yielding the desired number of states for MSM construction. For various conformational changes of biological molecular systems, previous studies elucidate their free energy landscape is hierarchical.\cite{huang2010constructing, yao2013hierarchical} A smaller $\Delta t$ may yield higher-resolution MSMs, identifying numerous free energy minima separated by small barriers as metastable states, whereas a larger $\Delta t$ may result in lower-resolution MSMs, detecting fewer metastable states encompassing multiple local free energy minima. SPIB provides great flexibility for users to select the most suitable resolution for MSMs to investigate various biological phenomena.

\revone{We note that, while the SPIB models shown here typically show good performance regarding the GMRQ score, the SPIB loss function (eq. \ref{eq:SPIB_obj}) is not equivalent to the GMRQ or VAMP-based scores, though Both VAMP-based scores and SPIB loss aim to approximate the propagator of underlying dynamics. The GMRQ score used in VAMP-based approaches relies on maximizing the sum of the top n eigenvalues of transition probability matrix constructed on the coarse-grained  subspace of the relevant, slow dynamics. In contrast, the SPIB loss function is the evidence lower bound to the maximum likelihood estimate of the state transition density. It comprises a reconstruction term and a regularization term. In the limit that the hyper-parameter $\beta$ governing the regularization term in the SPIB loss function approaches zero, the SPIB loss function focuses solely on minimizing the error of propagator approximation. Consequently, optimizing the SPIB loss in this scenario may yield outcomes akin to those derived from the GMRQ score. However, it's worth noting that a key distinction from VAMP-based methods remains: SPIB focuses on slow transitions with significant Boltzmann weights, whereas VAMP-based approaches concentrate solely on the slowest processes without considering their weights. This unique aspect allows SPIB to consistently learn highly populated metastable states, resulting in higher entropy scores in the systems tested in this study.}

\revone{In this work, to maintain consistency with prevailing practices in the MSM literature, we chose to conduct hyperparameter selection based on the GMRQ rather than the SPIB loss function. Nevertheless, recognizing both the connections and distinctions between the SPIB loss and VAMP-based scores, we acknowledge the potential of exploring the direct use of the SPIB loss for cross-validation as a valuable avenue for future research.}

\revone{The combination of the SPIB loss function, which incorporates a heuristic for state metastability, with the flexible use of lag time $\Delta$t allows for the construction of multi-resolution MSMs that perform well over the range of evaluation metrics provided in Tables \ref{tab:large_dt_quantitative_comparison} and \ref{tab:moderate_dt_quantitative_comparison}. Furthermore, the ability of the SPIB to learn adaptively the number of metastable states on-the-fly and as a function of lagtime differentiate the SPIB models from the analogous VMAPnet models.}

\revone{In addition, the SPIB consistently learns a low-dimensional latent space on-the-fly, effectively embedding the transitions between multiple metastable states of interest. This contrasts with existing dimension reduction methods, such as tICA and PCA, which typically fall short in capturing transitions between multiple metastable states within a 2D space (see SI).}

\revone{Overall, we believe this one-shot nature of the SPIB, where a latent space for description of the slow dynamics and kinetic modeling is built simultaneously, provides a low dimensional embedding of the metastable states of interest and is useful to analyze metastable dynamics in a variety of systems, including the protein folding dynamics analyzed here.}

\section{Conclusion}
In this work, we use the state predictive information bottleneck (SPIB)\cite{Wang2021} for constructing multi-resolution MSMs from MD simulation trajectories of protein conformational changes. The framework integrates the variational information bottleneck principle with a simple heuristic of the state metastability using a flexible neural network to simultaneously achieve feature extraction and state partitioning in a unified approach. By employing diverse quantitative and qualitative metrics across three distinct mini-proteins (Trp-cage, villin headpiece [HP35], and WW-domain), our study elucidates the distinct advantages of the SPIB approach over competing methods, all while requiring minimal hyperparameter tuning. These advantages include its ability to automatically determine the number of metastable states based on the specified minimum time resolution of interest, achieving a superior trade-off between capturing slow dynamics and states with significant population, and providing a direct visualization of underlying dynamics. 

Given the specified lag time $\Delta t$, which quantifies the minimal time resolution of interest, SPIB automatically adjusts the number of metastable states in the system. This eliminates the need to manually tune the number of output states, a step commonly required in other methods. Thus, SPIB offers users a straightforward approach to constructing MSMs for complex systems across diverse resolutions.

Without explicit optimization of the VAMP-based score, SPIB consistently demonstrates state-of-the-art performance in capturing the foremost slow dynamical processes, achieving comparable or slightly superior performance in validation GMRQ and metastability. Additionally, the top few slowest ITS of MSMs constructed by SPIB exhibit rapid convergence to their timescales even with shorter lag times implying an accurate Markovian kinetic model.

Beyond its proficiency in capturing slow dynamics, SPIB presents a distinct advantage when constructing a more nuanced MSM with 10 states. While VAMP-based methods optimize the overall kinetic performance of the model, they can struggle to further subdivide highly populated states. In contrast, SPIB excels in learning numerous well-populated macrostates. This capability stems from SPIB's optimization of likelihood through the information bottleneck principle, where only slow transitions with significant probabilities contribute. This characteristic sets SPIB apart from VAMP-based methods, allowing SPIB to excel in capturing intricate structural details. This efficacy enables effective differentiation among various metastable states, particularly in discerning subtle differences within folded or unfolded conformations in the study of protein folding processes.

Our results also indicates that SPIB learns a low-dimensional, continuous embedding of MD conformations that could preserve maximally the information about the state-to-state transitions. This capability facilitates direct visualization and provides a more insightful interpretation of the folding and unfolding pathways in mini-proteins through a 2D space. This stands in contrast to many existing dimension reduction methods which directly approximate eigenfunctions. 

Upon exploring all three mini-proteins, the findings from SPIB point towards a hierarchical organization in the free energy landscape governing their folding processes. This organizational structure involves the segmentation of both the native and unfolded basins into several well-populated metastable states, along with the existence of a few less populated intermediate or misfolded states. Even in the case of these seemingly simple proteins, the folding processes are multifaceted, encompassing multiple folding and unfolding pathways. Overall, we believe our algorithm introduces a novel, practical, and robust approach to construct MSMs, with potential applications across molecular simulations and the analysis of complex dynamical systems. We anticipate its utility across diverse scientific communities.

\section{Supporting information}
See supporting information for  a detailed description of model hyperparameters, a discussion on the impact of training hyperparameters and initial state labels, as well as additional supplementary results. \newline

\textbf{Acknowledgements\newline }
We thank D. E. Shaw Research for sharing the Trp-cage, HP35, and WW domain MD trajectories analyzed in this work. We thank UMD HPC’s Zaratan and NSF ACCESS (project CHE180027P) for computational resources. We acknowledge computational resource support from the Center for High Throughput Computing at the University of Wisconsin-Madison. We acknowledge the support by the NIH/NIGMS under award number 1R35GM142719 (P.T., D.W. and E.B.) the NIH/NIGMS under award number 1 R01GM147652-01A1 (X.H.) and the Hirschfelder Professorship Fund from University of Wisconsin-Madison (X.H.). P.T. was an Alfred P. Sloan Foundation fellow during the preparation of this manuscript, and is currently an investigator at the University of Maryland-Institute for Health Computing, which is supported by funding from Montgomery County, Maryland and The University of Maryland Strategic Partnership: MPowering the State, a formal collaboration between the University of Maryland, College Park and the University of Maryland, Baltimore. \newline

\textbf{Code availability statement\newline }
The SPIB codes for trajectory analysis can be accessed for public use on https://github.com/wangdedi1997/spib. PCCA+ clustering was conducted using the deeptime software package\cite{Hoffmann2022}, MPP clustering using the code from Gerhard Stock's group\cite{moldyn}, and other MSM calculations were performed using MSMBuilder\cite{Harrigan2017}.\newline

\textbf{References}
\bibliography{references}

\begin{figure*}[b!]
    \centering
    \includegraphics[width=1\textwidth]{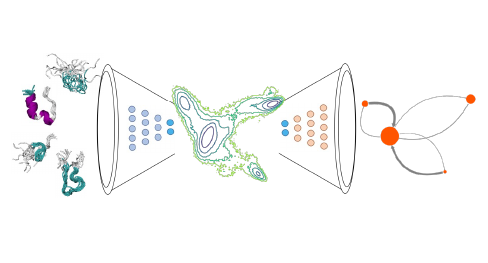}
    \caption{TOC.}
    \label{fig:toc}
\end{figure*}
\end{document}